\documentclass[aps, reprint,twocolumn,superscriptaddress]{revtex4-2}
\usepackage{amsmath}
\usepackage{latexsym}
\usepackage{amssymb}
\usepackage{graphics,epstopdf}
\usepackage{hyperref}
\usepackage{soul}
\usepackage{braket}
\hypersetup{
    colorlinks = true,
    linkcolor =blue,
	citecolor=blue, 
	urlcolor=blue 
}

\newcommand{\ed}{\end{document}}
\newcommand{\beq}{\begin{equation}}
\newcommand{\eeq}{\end{equation}}

\usepackage{mathtools}
\usepackage{natbib}

\newtheorem{lemma}{Lemma}
\newtheorem{theorem}[lemma]{Theorem}

\begin{document}
\title{Arbitrary order transfer matrix exceptional points and van Hove singularities}

\author{Madhumita Saha}
\email{madhumita.saha@icts.res.in} 
\affiliation{International Centre for Theoretical Sciences, Tata Institute of Fundamental Research,
Bangalore 560089, India}

\author{Bijay Kumar Agarwalla}
\email{bijay@iiserpune.ac.in}
\affiliation{Department of Physics, Indian Institute of Science Education and Research Pune, Dr. Homi Bhabha Road, Ward No. 8, NCL Colony, Pashan, Pune, Maharashtra 411008, India}

\author{Manas Kulkarni}
\email{manas.kulkarni@icts.res.in} 
\affiliation{International Centre for Theoretical Sciences, Tata Institute of Fundamental Research,
Bangalore 560089, India}

\author{Archak Purkayastha}
\email{archak.p@phy.iith.ac.in}
\affiliation{Department of Physics, Indian Institute of Technology, Hyderabad 502284, India}

\date{\today} 
\begin{abstract}
In lattice models with quadratic finite-range Hermitian Hamiltonians, the inherently non-Hermitian transfer matrix (TM) governs the band dispersion. The van Hove singularities (VHSs) are special points in the band dispersion where the density of states (DOS) diverge. Considering a lattice chain with hopping of a finite range $n$, we find a direct fundamental connection between VHSs and exceptional points (EPs) of TM, both of arbitrary order. In particular, we show that VHSs are EPs of TM of the same order, thereby connecting two different types of critical points usually studied in widely different branches of physics. Consequently,  several properties of band dispersion and VHSs can be analyzed in terms of spectral properties of TM. We further provide a general prescription to generate any order EP of the TM and therefore corresponding VHS. For a given range of hopping $n$, our analysis provides restrictions on allowed orders of EPs of TM. Finally, we exemplify all our results for the case  $n=3$.
\end{abstract}

\maketitle


\noindent

{\it Introduction} ---
Singularities play a crucial role across all branches of physics. They represent critical points of the underlying theory. Various universal and counterintuitive phenomena arise close to the critical points \cite{singularity_book,Sachdev_2011,yeomans1992statistical}. We find a fundamental connection between two types of critical points, which are usually discussed in widely different branches of physics. 

On one hand, we have the van Hove singularities (VHSs) in solid state physics \cite{van_hove_1953,Ashcroft,Mahan}. They are the critical points of band structure of a solid, and correspond to energies where the gradient of single-particle energy dispersion of the system vanishes.  The density of states (DOS) diverges at the VHSs. Close to these critical points of the single-particle description, many-body effects are expected to become crucial, leading to interesting emergent phenomena \cite{Shtyk_2017,1D_van_hove,Yuan_2019,Scammell_2D_van_hove,critical_topology_Liang,Efremov_2019,Classen_2020,Salamon_2020,Chadrasekaran_2020,Guerci_2022,Hu_2022,Chandrasekaran_2022,Shankar_2023,Sheffer_2023,Zakharov_2024}. If, in addition to the gradient, some higher order derivatives of the dispersion relation are also zero, they are termed higher order VHSs. Higher order VHSs have recently been of significant interest owing to their occurrence in van der Waals materials, where they have been associated with various observed exotic properties \cite{Yuan_2019,Chandrasekaran_2022,Guerci_2022, Shankar_2023}. 

On the other hand, we have the exceptional points (EPs) in non-Hermitian physics \cite{Chen_2022, Abbasi_2022, Liao_2021, Liu_2021, Chen_2020, Naghiloo_2019, Bergholtz_2021,Ozdemir_2019, El_Ganainy_2018, Feng_2017,Longhi_2017, Konotop_2016}. They are critical points in parameter space where a non-Hermitian matrix, appearing in the physical description of a system, becomes non-diagonalizable. They occur when several eigenvalues and corresponding eigenvectors of the matrix coalesce. The number of coalescing eigenvectors  give the order of the EP. Having no analog in Hermitian matrices, EPs are extensively studied theoretically and experimentally, in classical and quantum systems \cite{non-hermitian-review, Bergholtz_2021,Ozdemir_2019, El_Ganainy_2018, Feng_2017,Longhi_2017, Konotop_2016}. Most commonly studied EPs are of second order, while engineering hypersurfaces of higher order EPs have been of more recent interest \cite{general_hatano_nelson, higher_order_exceptional,open_exceptional,toplology_non_hermitian,EPD_photonic,wave_guideEPD,symmetry_higher_order_EP,Bergholtz_2021,finite-range-subdiffusive,saha2024effect}. Occurrence of EPs have also found various applications, including devising sensors, lasers and perfect absorbers \cite{sensing_EPD,sensing1,sensing_review,sensing_review1,band_edge_EP_transfer_matrix,spontaneous_emission,perfect_absorber,non-hermitian-review, Bergholtz_2021,Ozdemir_2019, El_Ganainy_2018, Feng_2017,Longhi_2017, Konotop_2016}.

In this Letter, considering a chain with finite-range hopping, we provide a deep connection between VHSs and EPs. The energy dispersion of the translationally invariant Hermitian Hamiltonain in the thermodynamic limit can be obtained in terms of a finite-size transfer matrix (TM), which is non-Hermitian by construction. We show that the VHSs correspond to the EPs of the TM, of the same order. Consequently, properties of band dispersion of the Hermitian system can be analyzed in terms of the spectral properties of the non-Hermitian TM. We provide a general way to generate arbitrary order EP of TM, and thereby VHS of same order, by controlling the hopping parameters and the range of hopping. For a given range of hopping, we provide restrictions on the allowed orders of EPs of TM.  We discuss the subtle differences between VHSs occurring at band extrema and those at saddle points, and demonstrate an explicit example.

{\it The system, energy dispersion, and TM ---} We consider an isolated system of a one-dimensional lattice with hopping of finite-range $n$. The Hamiltonian can be written as
$
\hat{H}=-\sum_{m=1}^n \sum_{i=-\infty}^{\infty} t_m \hat{c}^{\dagger}_i \hat{c}_{i+m} + h.c.,
$
where, $\hat{c}^{\dagger}_i$ $(\hat{c}_i)$ is the fermionic or bosonic creation (annihilation) operator, and $h.c.$ stands for Hermitian conjugate. Here, $t_m$ is the hopping strength when the particle hops from $i$th site to $(i+m)$th site. The dispersion relation for this system can be obtained by diagonalizing single particle Hamiltonian and is given by
$\epsilon(k)=-2\sum_{m=1}^n t_m \cos(mk)$,
where $k$ is the lattice Bloch momentum vector and ranges between $-\pi \leq k \leq \pi$, i.e, within the Brillouin zone. The range of possible values of $\epsilon(k)$, for $k$ within the Brillouin zone, defines the system band. The density of states (DOS) of the system is given in terms of the dispersion relation as
$
\nu(\omega)=\int_{-\pi}^\pi \frac{dk}{2\pi} \, \delta[\omega-\epsilon(k)], 
$
where $\delta(x)$ is the Dirac delta function. The VHSs are values of $\omega$ where $\nu(\omega)$ diverges \cite{van_hove_1953, Ashcroft,Mahan}. They occur at energies corresponding to extrema and saddle points of the band dispersion, i.e, where one or more derivatives of $\epsilon(k)$ are zero. Such points are also called the critical points of the system band. Let us define 
\begin{align}
\label{a_r}
a_r(k_0)=\frac{d^r \epsilon(z)}{d z^r}\Big |_{z=k_0}\hspace*{-12pt}=\,\hspace*{-1pt}-2 \hspace*{-3pt}\sum_{m=1}^n \hspace*{-3pt} t_m m^r \cos\left(\frac{r \pi}{2}+m k_0\right).
\end{align}   
If $a_r(k_0)=0$ for $r=1,2,\ldots,p-1$, and $a_p(k_0)\neq 0$, then $\epsilon(k_0)$ is said to be a $p$th order critical point. The smallest order of critical point is therefore $p=2$. Critical points of order $p>2$ are called higher order VHSs. In what follows, we give an interesting connection between the  VHSs and EPs of the TM of the lattice system.

The TM of this system is obtained by writing the discrete Schr\"odinger equation for the single particle eigenvectors with energy $\omega$ \cite{matrix_inversion,finite-range-subdiffusive,Molinari_1997,Molinari_1998,transfer_matrix1,transfer_matrix2}.
The TM $\mathbf{T}(\omega)$ is a $2n \times 2n$ dimensional non-Hermitian matrix with real entries, where, we remind that $n$ is the range of hopping. Its determinant is $1$. The non-zero elements of the TM are: $\mathbf{T}_{j+1~j}(\omega)=1,~~\forall~j=1,2,\ldots 2n-1$; $\mathbf{T}_{1~j}(\omega)=-\frac{t_{n-j}}{t_n}~\forall~j=1,2,\ldots, n-1$; $\mathbf{T}_{1~n}(\omega)=-\frac{\omega}{t_n}$; $\mathbf{T}_{1~j}(\omega)=-\frac{t_{j-n}}{t_n},~\forall~j=n+1,n+2,\ldots,2n$. The rest of the elements are zero. 
By virtue of having real entries, $\mathbf{T}(\omega)$ has a trivial anti-unitary symmetry $\hat{\mathcal{K}}\mathbf{T}(\omega)\hat{\mathcal{K}}^{-1}=\mathbf{T}(\omega)$, where $\hat{\mathcal{K}}$ is the complex conjugation operator,  $\hat{\mathcal{K}}M\hat{\mathcal{K}}^{-1}=M^*$. This guarantees that eigenvalues of $\mathbf{T}(\omega)$ are either real or comes in complex conjugate pairs \cite{non-hermitian-review}. The connection between the $\epsilon(k)$ and $\mathbf{T}(\omega)$ is given by the fact that the eigenvalues and eigenvectors of $\mathbf{T}(\omega)$ are obtained as
\begin{equation}
\label{T_eigvals_eigvecs_relation}
    \mathbf{T}(\omega)\ket{\phi^{(n)}(z)}=e^{-iz}\ket{\phi^{(n)}(z)},\,\, \forall~~\omega=\epsilon(z).
\end{equation}
where $\ket{\phi^{(n)}(z)}$ is a column vector with $2n$ elements, whose  $j$th element is given by $e^{-iz(n-j+1)}$. 
In Eq.~\eqref{T_eigvals_eigvecs_relation}, $z$ can be complex, and we have analytically continued the dispersion relation to complex arguments. Defining $x=\cos z$, we see that, for given $\omega$, $\omega= \epsilon(z)$ in Eq.~\eqref{T_eigvals_eigvecs_relation} becomes $\omega~+~2\sum_{m=1}^n t_m \mathbb{T}_m(x)~=~0$, where $\mathbb{T}_m(x)$ is the $m$th Chebyshev polynomial of first kind.  The left-hand-side is a polynomial of degree $n$ in $x$, and therefore can be solved for $n$ roots, thereby obtaining $n$ corresponding values of $z$. Further, since $\epsilon(z)$ is an even function, if $z$ satisfies $\omega=\epsilon(z)$ in Eq.~\eqref{T_eigvals_eigvecs_relation}, so does $-z$. Thus, the $n$ roots in terms of $x$ gives $2n$ values of $z$ satisfying Eq.~\eqref{T_eigvals_eigvecs_relation}. These $2n$ values of $z$ specifies all the $2n$ eigenvalues and eigenvectors of $\mathbf{T}(\omega)$ via Eq.~\eqref{T_eigvals_eigvecs_relation}.

For the solutions of $x$ in the regime, $-1\leq x \leq 1$, $z$ is real, and can be taken to be in the Brillouin zone, thereby corresponding to the lattice Bloch momentum vector $k$. From Eq.~\eqref{T_eigvals_eigvecs_relation}, the corresponding eigenvalues of $\mathbf{T}(\omega)$ are of unit modulus.
Such eigenvalues can exist only when $\omega$ is chosen to be within the system band. The number of eigenvalues  of $\mathbf{T}(\omega)$ with unit modulus gives the number of values of $k$ corresponding to $\omega$.
\\

{\it EPs of $\mathbf{T}(\omega)$ and VHSs---} For $\mathbf{T}(\omega)$, we see from Eq.~\eqref{T_eigvals_eigvecs_relation} that several eigenvalues and eigenvectors can coalesce if and only if solving $\omega=\epsilon(z)$ for $z$ yields repeated roots. Furthermore, noting that $\epsilon(z)$ is an analytic function of $z$, we have the following theorem (see Appendix.~\ref{appendix1}).
\begin{theorem}
	\label{EP_theorem}
For a given $\omega=\omega_0$, the TM $\mathbf{T}(\omega_0)$ has an EP of order $p$, iff, there exists $k_0$ such that (i) $\omega_0=\epsilon(k_0)$, (ii) $a_r(k_0)=0$, $\forall~r < p$, and (iii) $a_p(k_0)\neq 0$.
\end{theorem}
In above, $a_r(k_0)$ is defined in Eq.~\eqref{a_r}, and $k_0$ is complex in general. If $k_0$ is real, then Theorem~\ref{EP_theorem} says that $p$th order EP of TM corresponds to $p$th order critical point of the system band. Odd order EPs of $\mathbf{T}(\omega)$ correspond to saddle points, even order EPs of $\mathbf{T}(\omega)$ correspond to extrema (maxima if  $a_p(k_0)< 0$, minima if $a_p(k_0)> 0$). Thus, there is a fundamental connection between EPs of $\mathbf{T}(\omega)$  and VHSs. 
This connection is made more explicit via our following theorem, which straightforwardly follows from application of Theorem \ref{EP_theorem}. 
\begin{theorem}
\label{VHS_theorem}
Let $\omega_0$ be such that some unit modulus eigenvalues of $\mathbf{T}(\omega_0)$ are repeated $p$ number of times, $p>1$, i.e, $\omega_0$ is within system band and is the $p$th order EP of TM. Let $c(\omega)$ be the number of unit modulus eigenvalues of $\mathbf{T}(\omega)$. Define the index $\mathcal{I}~=~{\rm sign}\big[c(\omega_0~+~\delta)-c(\omega_0-\delta)\big]$, with $\delta>0$.  The following statements hold:
\begin{itemize}
\item[(i)] Allowed values of order of EP for a given range of hopping $n$-- 

\noindent (a) For a given range of hopping $n$, the allowed values of $p$ are all integers from $2$ to $n$, and all even integers between $n$ and $2n$. \\  

\noindent (b) EPs of order $p>n$ can only occur at $k=0,\pm \pi$.

\item[(ii)] Relationship between divergence of DOS and order of EP--

\noindent (a) Iff $p$ is even and $\mathcal{I}=1$, then the DOS $\nu(\omega_0+\delta)\propto \delta^{-(p-1)/p}$. This corresponds to a minimum of energy dispersion.
\\
\\
\noindent (b) Iff $p$ is even and $\mathcal{I}=-1$, then $\nu(\omega_0-\delta)\propto \delta^{-(p-1)/p}$. This corresponds to a maximum of energy dispersion.\\ \\
\noindent (c) Iff $p$ is odd, then, $\mathcal{I}=0$, and $\nu(\omega_0~\pm \delta)\propto \delta^{-(p-1)/p}$. This corresponds to a saddle point of energy dispersion.
\end{itemize}
\end{theorem}
Theorems \ref{EP_theorem} and \ref{VHS_theorem} are the central results of this Letter. Part (i) of Theorem \ref{VHS_theorem} puts restrictions on the allowed orders of EPs of TM for a given range of hopping.
Part (ii) of Theorem \ref{VHS_theorem} shows how the nature of VHS, i.e, whether it corresponds to band minimum, maximum or saddle point, as well as the exponent with which the DOS diverges, can be found from the spectral properties of TM. Interestingly, the index $\mathcal{I}$ defined in terms of the TM eigenvalues, can also be written as $\mathcal{I}=[1+(-1)^p]\,{\rm sign}[a_p(k_0)]/2$. This is nothing but the topological index defined in Ref.~\onlinecite{critical_topology_Liang}  while phenomenologically discussing higher-order VHSs in one-dimension. Following Ref. \onlinecite{critical_topology_Liang}, physically $\mathcal{I}=1$ ($\mathcal{I}=-1$), which corresponds to band minimum (maximum), means that the effective velocity of particles changes sign from negative (positive) to positive (negative) on crossing the critical point. Contrarily, $\mathcal{I}=0$, which corresponds to saddle point, means that there is no change in direction of effective velocity of particles on crossing the critical point. Also note that, divergence of DOS occurs on approaching a band minimum (maximum) from above (below), while divergence occurs on approaching a saddle point from either side.
\\

Next, we outline the proof of Theorem \ref{VHS_theorem}, which gives further insights. We start by proving part (i) of Theorem \ref{VHS_theorem}. The proof is given by using Theorem \ref{EP_theorem} (see Appendix.~\ref{appendix1}) to formulate a general prescription for obtaining arbitrary order exceptional hypersurfaces of the TM by varying the hopping strengths  and the range of hopping of the lattice.

{\it Generating arbitrary order exceptional hypersurfaces of $\mathbf{T}(\omega)$---}
Following Theorem \ref{EP_theorem}, finding conditions for an arbitrary order EP of TM requires equating several derivatives of dispersion relation to zero. From Eq.~\eqref{a_r}, we readily see that $k=0, \pm \pi$, all odd derivatives become zero. So, irrespective of the values of hopping strengths $t_m$, the TM has at least a second order EP at $k=0, \pm \pi$. The corresponding values of $\omega$ are
$\omega=\epsilon(0)=-2 \sum_{m=1}^n t_m,~~{\rm for}~ k=0$, $\omega=\epsilon(\pm \pi)=-2 \sum_{m=1}^n (-1)^m t_m,~~{\rm for}~ k=\pm \pi.$
We consider $k=0$ case. Similar discussion holds for $k=\pm \pi$ case.
Let us set the hopping parameter $t_1=1$, which therefore sets the energy scale. In the $n$ dimensional parameter space which is spanned by $\omega$, and the remaining $n-1$ hopping parameters, $\omega =\epsilon(0)$ defines a $n-1$ dimensional hypersurface, all points on which are EPs of the TM, of at least second order.

Next, keeping $k=0$, we can sequentially put $r=2,4,6,8, \ldots, 2n-2$ in Eq.~\eqref{a_r} and equate them to zero, to find more and more constraints on the allowed values of $t_2, t_3, \cdots t_{n}$. Such increasing number of constraints sequentially defines lower dimensional hypersurfaces in the parameter space, every point on which is a higher order EP of the TM. Finally, setting all the expressions for $r=2,4,\ldots 2n-2$ in Eq.~\eqref{a_r} to zero, there are $n-1$ constraints. Thus all $n-1$ parameters $t_2, t_3, \cdots t_{n}$ are uniquely defined. This leads to a single point in the parameter space corresponding to $2n$th order EP. Since TM is a $2n \times 2n$ dimensional matrix, this is the highest order EP possible. In this way, we can systematically generate any even order EP of the TM.

In contrast, odd order EPs are not possible for $k=0,\pm \pi$, because all odd derivatives of dispersion relation are zero. A $p$th order EP of TM for $k=k_0\neq 0,\pm \pi$ can be generated similarly as above by sequentially equating all derivatives up to order $p-1$ to zero. But, since $\epsilon(k)$ is an even function of $k$, if $e^{-ik_0}$ is a degenerate eigenvalue of $\mathbf{T}(\omega)$, then $e^{ik_0}$ is also a different eigenvalue with the same degeneracy. Therefore, all such EPs occurring at $k=k_0\neq 0,\pm \pi$ occur in pairs. Consequently, since the TM is a $2n\times 2n$ matrix, the highest order for EPs occuring away from $k=0,\pm \pi$ is $n$.  Combining this with our previous discussion of even order EPs provides the proof of part (i) of Theorem \ref{VHS_theorem}. Next we consider the proof of part (ii) of  Theorem \ref{VHS_theorem}, which is based on analyzing the non-analytic change that occurs in spectrum of TM across an EP.

{\it Transition across EP---} Let there be a $p$th order EP of TM at $\omega=\omega_0=\epsilon(k_0)$, $k_0$ being real. Let $\omega_0\pm\delta = \epsilon(z_{\pm}(k_0))$, with $\delta>0$. After some algebra, using Theorem~\ref{EP_theorem}, $z_{\pm}(k_0)$ is given to the leading order in $\delta$ as
\begin{align}
\label{z_pm}
&z_{\pm}(k_0) =
\begin{cases}
& \hspace*{-10pt} k_0 + B_p(k_0) e^{i2m\pi/p},~~ {\rm if}~~{a_p(k_0)} \gtrless 0  \\
& \hspace*{-12pt} k_0 + B_p(k_0) e^{i(2m+1)\pi/p}, ~~ {\rm if }~~{a_p(k_0)} \lessgtr 0 
\end{cases}
\end{align}
where $B_p(k_0)=\left(\frac{p! \delta}{|a_p(k_0)|}\right)^{1/p}$ and $m=0,1,2,\ldots,p-1$. This expression shows that on infinitesimal change of $\omega$ from $\omega_0$, $k_0$ splits into $p$ different values, many of them being complex. This should be contrasted against the case away from EP, which can be analyzed by putting $p=1$ in Eq.~\eqref{z_pm}. Away from EP, there is only one value of $z_{\pm}(k_0)$, and the value remains real. The connection between EPs of TM and VHSs depends on the real values of $z_{\pm}(k_0)$, which, from Eqs.~\eqref{T_eigvals_eigvecs_relation}, corresponds to unit modulus eigenvalues of TM.

First consider $p$ being even, i.e, $p=2s$ and $a_{2s}(k_0)>0$ (band minima from Theorem~\ref{EP_theorem}). Substituting in Eq.~\eqref{z_pm}, we see that $z_{+}(k_0)$ has two real values, corresponding to $m=0$ and $m=s$, given by $k_0+B_{p}(k_0)$ and $k_0-B_{p}(k_0)$ respectively. On the contrary, there are no purely real values of $z_{-}(k_0)$. Thus, the number of unit modulus eigenvalues of TM increases on crossing the EP from below. In other words, $c(\omega_0 + \delta)>c(\omega_0 - \delta)$, where $c(\omega)$ is defined in Theorem~\ref{VHS_theorem}. Moreover, since there are no purely real $z_{-}(k_0)$, and DOS is only affected by real $k$ values, it follows that $\nu(\omega_0 - \delta)$ is unaffected by the EP. But, real values of $z_{+}(k_0)$ affect $\nu(\omega_0 + \delta)$, and using the expression for $B_p(k_0)$, it can be shown that $\nu(\omega_0+\delta)=A_p(k_0) \delta^{-(p-1)/p}$, where $A_p(k_0)=\frac{(p-1)!}{\pi |a_p(k_0)| }\left[\frac{|a_p(k_0)|}{p!}\right]^{(p-1)/p}$ (see Appendix.~\ref{app2}). 

If $p=2s$ and $a_{p}(k_0)<0$ (band maxima from Theorem~\ref{EP_theorem}), from Eq.~\eqref{z_pm}, we get that there are two real values of $z_{-}(k_0)$, and no purely real values of $z_{+}(k_0)$. It therefore follows that, $c(\omega_0 + \delta)<c(\omega_0 - \delta)$ and $\nu(\omega_0 + \delta)$ is unaffected by the EP, while $\nu(\omega_0-\delta)=A_p(k_0) \delta^{-(p-1)/p}$.

Finally consider $p$ being odd, i.e, $p=2s+1$ (saddle point of band from Theorem~\ref{EP_theorem}). We assume $a_{2s+1}(k_0)>0$ in the following, similar discussion is possible for $a_{2s+1}(k_0)<0$. Substituting in Eq.~\eqref{z_pm} we see that in this case, $z_{+}(k_0)$ has one real value, corresponding to $m=0$, given by $k_0+B_p(k_0)$, while $z_{-}(k_0)$ also has one real value, corresponding to $m=s$, given by $k_0-B_p(k_0)$. Thus, the number of unit modulus eigenvalues of TM does not change on crossing the EP,  $c(\omega_0 + \delta)=c(\omega_0 - \delta)$. Moreover, since there is one real value of both $z_{+}(k_0)$ and $z_{-}(k_0)$, DOS is similarly affected by the EP on approaching from either side, giving $\nu(\omega_0\pm\delta)=A_p(k_0) \delta^{-(p-1)/p}$. The above completes the proof of all the statements in Theorem~\ref{VHS_theorem}(ii). In the following, we demonstrate $n=3$ case as an explicit example.

\begin{figure}
\includegraphics[width=\columnwidth]{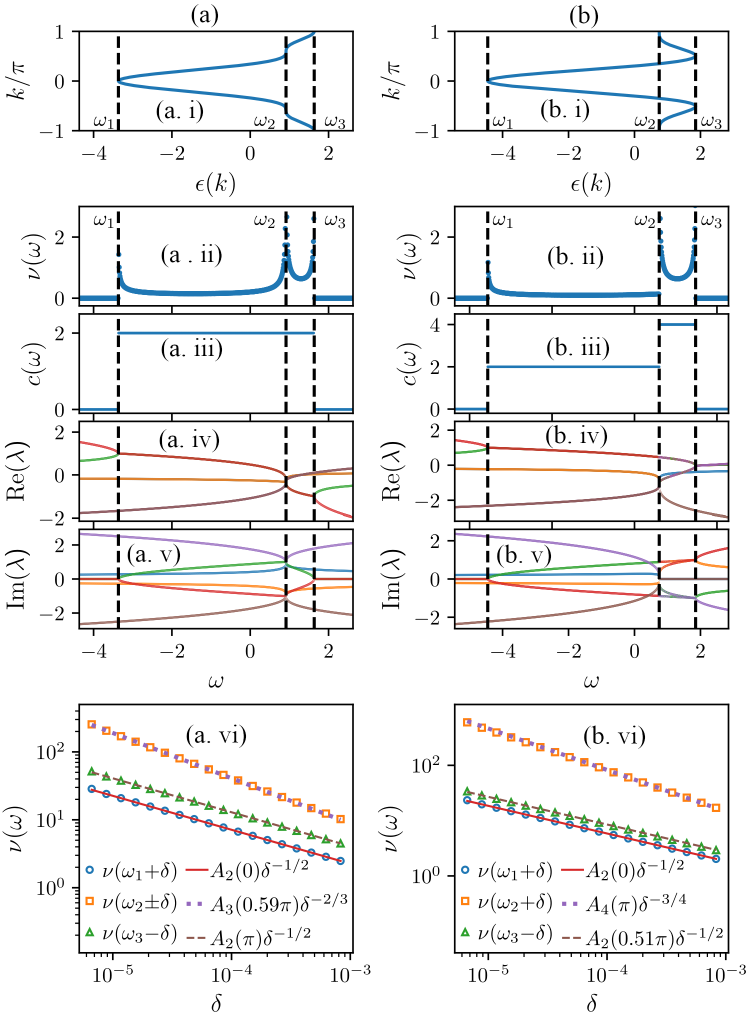}
\caption{\label{fig:plots} Column (a) corresponds to $t_1=1$, $t_2=\sqrt{3}/{4}$, $t_3=1/4$,  condition to host a third order EP at $k=k_{*}=\cos^{-1} \big[-1/\sqrt{12}\big]$. Column (b) corresponds to $t_1=1$, $t_2=37/40$, $t_3=3/10$,  condition for a fourth order EP at $k=\pi$. The vertical dashed lines in rows (i)-(v) for both columns show the energies corresponding to the critical points: $\omega_1$, $\omega_2$, $\omega_3$. In rows (iv) and (v) of both columns, ${\rm Re}(\lambda)$ and  ${\rm Im}(\lambda)$ represents the real and imaginary parts of the eigenvalues of TM, respectively.
Each color represents a different eigenvalue. Both real and imaginary parts of each eigenvalue have the same color.  
In row (vi) of both columns, we show the nature of DOS in the vicinity of EPs of TM. The symbols here represent numerically obtained DOS, while the dashed, dotted and continuous lines are plots obtained via analytical result $A_p(k_0) \delta^{-(p-1)/p}$, where $p$ is the order of EP of TM at $\omega=\epsilon(k_0)$. All energies are in units of $t_1$ and we take $n=3$.}
\end{figure}

{\it Illustrative example, $n=3$ ---} The parameter space with the lattice of range of hopping $n=3$ is three-dimensional consisting of $(\omega,t_2,t_3)$, with $t_1=1$ being chosen as the unit of energy. The critical points  $k$ of the band dispersion can be found by solving $a_1(k)=0$. This yields, $k=k_0=0,\pm \pi, k_*$, with $k_*=\cos^{-1}\Big[\frac{1}{2}\big{(}-\frac{t_2}{3 \,t_3} \pm \sqrt{\big(\frac{t_2}{3\, t_3}\big)^2-\frac{t_1}{3 \,t_3}+1} \big{)}\Big]$. In the $(\omega,t_2,t_3)$ plane, $\omega=\epsilon(k_0)$ describes a surface each point on which is at least a second order EP of the TM. Now, choosing $k_0=\pi$, and setting $a_2(\pi)=0$ yields $t_2=t_1/4 + 9 t_3/4$. This gives a line on the $\omega=\epsilon(\pi)$ surface, each point on which is at least fourth order EP of the TM, since $a_3(\pi)$ is also zero. Next, setting $a_4(\pi)=0$ yields $t_2=2t_1/5, t_3=t_1/15$, which is a point on the line and corresponds to the sixth order EP, since $a_5(\pi)$ is also zero. This is the EP of highest order. Thus, we have all possible even order EPs at $k_0=\pi$.

For odd order EP, we take $k_0=k_*$ and set $a_2(k_*)=0$. This, along with the fact that $k_*$ and the hopping parameters are real yields the following condition for third order EP of TM:  $t_2^2/t_3^2 = 3 \, t_1/t_3-9$, with $t_1/15< t_3 < t_1/3$. For $n=3$, this is the highest order EP possible for $k_0\neq 0,\pm \pi$, as also follows from Theorem \ref{VHS_theorem} (i) (a) and (b).

In Fig.~\ref{fig:plots} columns (a) and (b), we show plots for a case where the hopping parameters satisfy the condition for a third order and a fourth order EP of TM, respectively. The energy dispersion plotted in Fig.~\ref{fig:plots}(a.~i), clearly shows saddle points, in addition to the global extrema (band edges), while that in Fig.~\ref{fig:plots}(b.~i) shows additional minima apart from the global extrema. Figures~\ref{fig:plots}(a.~ii), (b.~ii) show the numerically obtained DOS. We clearly see that the DOS diverges on approaching the minimum (maximum) from above (below), while it diverges on approaching the saddle point from either side. Figures~\ref{fig:plots} (a.~iii), (b.~iii) show $c(\omega)$, the number of unit modulus eigenvalues of $\mathbf{T}(\omega)$. They demonstrate that indeed $c(\omega)$ increases (decreases) on crossing a minimum (maximum) from below, while it remains unaffected by saddle point. Figures~\ref{fig:plots} (a.~iv), (b.~iv) and (a.~v), (b.~v) show the real and the imaginary parts of the eigenvalues of $\mathbf{T}(\omega)$. In Figs.~\ref{fig:plots}(a.~iv), (a.~v), we see that at the global minimum and maximum, which occurs at $k=0$ and $k=\pm \pi$, respectively, two eigenvalues merge, showing they are second order EPs. Contrarily, there are two sets of eigenvalues of TM, which merge at the saddle point, each set having three eigenvalues. This shows that there are a pair of third order EPs coexisting at the saddle point.  In Figures.~\ref{fig:plots} (b.~iv), (b.~v), we see that at the global minimum, which occurs at $k=0$, two eigenvalues merge, showing it is a second order EP of TM. At the local minimum occurring at $k=\pm \pi$, four eigenvalues merge, showing it is a fourth order EP. At the global maxima, which occurs at $k\neq 0,\pm \pi$, two sets of eigenvalues merge, each set having two eigenvalues. Thus, there are a pair of second order EPs of TM coexisting in this case. 

Finally, in Fig.~\ref{fig:plots} (a.~vi) and (b.~vi), we show the DOS on approaching the minima (maxima) from above (below) and saddle points from either side. We clearly see prefect matching between the numerically obtained DOS, and analytical prediction showing divergence as $\delta^{-(p-1)/p}$, where $p$ is the order of EP of TM. Therefore, Fig.~\ref{fig:plots} fully illustrates our previous general analytical discussion relating EPs of TM to VHSs of the same order.

{\it Summary and outlook ---}
In conclusion, we have established a fundamental connection between EPs, which are critical points in non-Hermitian physics, and VHSs, which are critical points of band structure of a solid. We highlight that unlike most works on non-Hermitian physics, we are not considering a dissipative system with any effective non-Hermitian description. Rather,  we only consider translationally invariant isolated Hermitian Hamiltonians in the thermodynamic limit. Even in such cases, the TM is non-Hermitian, and its EPs correspond to VHSs of the band structure. Our results hold for bosonic and fermionic systems alike.

More generally, for finite-ranged systems, the TM governs the system's Green's functions, thereby affecting correlations in the system \cite{universal_subdiffusive, superballistic1, finite-range-subdiffusive, hu_2024}. The  DOS, which shows the VHSs, is the imaginary part of the Green's function. Future work includes exploring the consequences of TM EPs on scatterring theory and diagrammatic perturbation techniques for many-body interactions. Going further, extending our results to two-dimensional systems will open the way for understanding the VHSs of van der Waals materials \cite{Yuan_2019,Chandrasekaran_2022,Guerci_2022, Shankar_2023} in terms of the non-Hermitian properties of the underlying TM, thereby connecting two exciting and active fields.

{\it Acknowledgements:}
 M. S. and M. K. acknowledge support of the Department of Atomic Energy, Government of India, under project no. RTI4001. B. K. A. would also like to acknowledge funding from the National Mission on Interdisciplinary  Cyber-Physical  Systems (NM-ICPS)  of the Department of Science and Technology,  Govt.~of  India through the I-HUB  Quantum  Technology  Foundation, Pune, India. B.K.A. acknowledges CRG Grant No. CRG/2023/003377 from SERB, Government of India.
M. K. thanks the VAJRA faculty scheme (No. VJR/2019/000079) from the Science and Engineering Research Board (SERB), Department of Science and Technology, Government of India.  M.K. acknowledges support from the Infosys Foundation International Exchange Program at ICTS. A.P acknowledges funding from Seed Grant from IIT Hyderabad, Project No. SG/IITH/F331/2023-24/SG-169. A.P also acknowledges funding from Japan International Coorperation Agency (JICA) Friendship 2.0 Research Grant, and from Finnish Indian Consortia for Research and Education (FICORE). M. K. thanks the hospitality of Department of Physics, IISER, Pune. B. K. A. and M. K. also thanks Katha Ganguly for insightful discussions related to the project.

\appendix

\setcounter{figure}{0}
\renewcommand{\thefigure}{A\arabic{figure}}

\section{Proof of Theorem 1 of the main text}
\label{appendix1}
\label{Theorem1}
In this section, we provide a proof for theorem 1 of the main text, which we restate below:

\setcounter{lemma}{0}
\begin{theorem}
	\label{EP_theorem_sup}
For a given $\omega=\omega_0$, the transfer matrix (TM) $\mathbf{T}(\omega_0)$ has an exceptional point (EP) of order $p$, if and only if, there exists $k_0$ such that (i) $\omega_0=\epsilon(k_0)$, (ii) $a_r(k_0)=0$, $\forall~r < p$, and (iii) $a_p(k_0)\neq 0$, where $a_r(k_0)=\frac{d^r \epsilon(z)}{d z^r}\Big |_{z=k_0}$.
\end{theorem}

{\it Proof--} 
The eigenvalue equation for $\mathbf{T}(\omega)$ is
\begin{equation}
\label{T_eigvals_eigvecs_relation_sup}
    \mathbf{T}(\omega)\ket{\phi^{(n)}(z)}=e^{-iz}\ket{\phi^{(n)}(z)},\,\, \forall~~\omega=\epsilon(z),
\end{equation} 
where $\ket{\phi^{(n)}(z)}$ is a column vector with $2n$ elements, whose  $j$th element is given by $e^{-iz(n-j+1)}$, and
\begin{equation}
\label{dispersion-supp}
   \epsilon(z) = -2 \sum_{m=1}^{n} t_m \cos(mz). 
\end{equation} 
So, to find the eigenvalues, we need to solve the following equation
 \begin{align}
\label{omega_epsilon_k_sup}
\omega-\epsilon(z)=0.
\end{align}
Here $z$ is complex in general.  
As mentioned in main text, there will be $2n$ solutions to Eq.~\eqref{omega_epsilon_k_sup}. Let them be denoted by $\{k_0,k_1,k_2,\ldots,k_{2n-1}\}$. Then,
the eigenvalues  of $\mathbf{T}(\omega)$ are given by $\{ e^{- i k_0}, e^{-i k_1}, e^{-i k_2}, \ldots, e^{-i k_{2n-1}} \}$. 

Let us say one solution, $k_0$, is known, i.e, $\omega=\epsilon(k_0)$. Using this, and the fact that $\epsilon(z)$ is an analytic function of $z$, we may obtain the other $z$ values satisfying Eq.~\eqref{omega_epsilon_k_sup} as follows. First, we expand $\epsilon(k)$ about $z=k_0$,
\begin{align}
\epsilon(z)=\epsilon(k_0) + \sum_{r=1}^{\infty} \frac{(z-k_0)^r}{r!} a_r(k_0),
\end{align}
where $a_r(k_0)$ is as given in Theorem \ref{EP_theorem_sup}. Then, to find other values of $z$ satisfying Eq.~\eqref{omega_epsilon_k_sup}, we impose $\omega=\epsilon(z)=\epsilon(k_0)$, to obtain
\begin{align}
\label{k_equation}
\sum_{r=1}^{\infty} \frac{(z-k_0)^r}{r!} a_r(k_0)=0.
\end{align}
The solutions of Eq.~\eqref{k_equation} gives the values of $z$ satisfying the dispersion Eq.~\eqref{omega_epsilon_k_sup}.   Now let us consider a situation where the first $p-1$ derivatives of $\epsilon(z)$ at $z=k_0$ vanishes, and the $p$th derivative does not vanish,
\begin{align}
\label{derivative_condition}
a_r(k_0)=0,~~\forall~1\leq r < p,~~~a_p(k_0) \neq 0.
\end{align}
We now prove that this is the necessary and sufficient condition for having a $p$th order EP of $\mathbf{T}(\omega)$.

{\it Sufficient ---}
If Eq.~\eqref{derivative_condition} holds, Eq.~\eqref{k_equation} becomes
\begin{align}
\label{p_repeated}
(z-k_0)^p \sum_{r=p}^\infty\frac{(z-k_0)^{r-p}}{r!} a_r(k_0)=0.
\end{align}
From Eq.~\eqref{p_repeated}, we see that $p$ solutions of Eq.~\eqref{k_equation} are the same, i.e, $k_0=k_1=k_2=\ldots=k_{p-1}$. Consequently, from Eq.~\eqref{T_eigvals_eigvecs_relation_sup} the $p$ eigenvalues and eigenvectors of $\mathbf{T}(\omega)$ coalesce.  In other words, the transfer matrix $\mathbf{T}(\omega)$ has $p$th order EP.  

{\it Necessary---}
Assume $\mathbf{T}(\omega)$ has $p$th order EP. From Eq.~\eqref{T_eigvals_eigvecs_relation_sup}, we see that this can only occur if solving Eq.~\eqref{omega_epsilon_k_sup} for $z$ yields $p$ repeated roots. This means Eq.\eqref{k_equation} should have $p$ repeated solutions when solved for $z$. This can only happen if Eq.~\eqref{derivative_condition} holds. 
This completes the proof of Theorem \ref{EP_theorem_sup}.



\begin{figure}
\includegraphics[width=0.7\columnwidth]{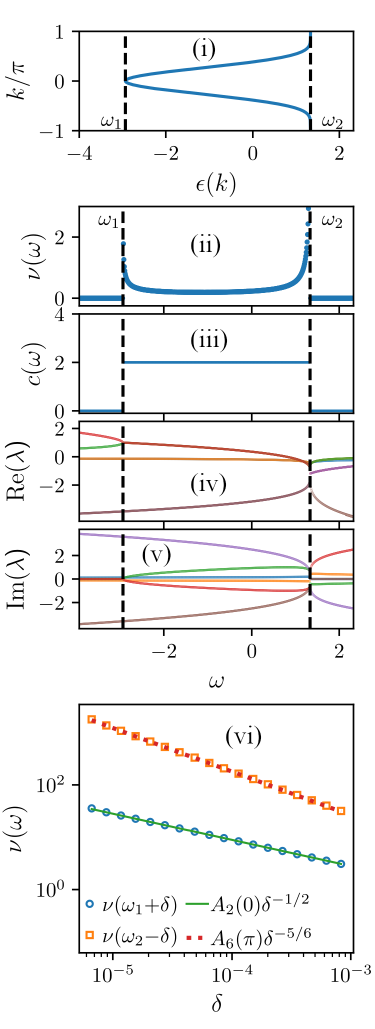}
\caption{\label{fig:sixth_order} The figure provides results in presence of a sixth order EP at  $k=\pi$ for $n=3$. 
We set $t_1=1$, $t_2=2t_1/5$, $t_3=t_1/5$, which is the condition to realize sixth order EP at $k=\pi$. Panel (i) shows the dispersion relation. Panel (ii) shows the DOS as a function of $\omega$. Panel (iii) shows the number of eigenvalues of unit modulus as a function of $\omega$. Panel (iv) and (v) shows the real and the imaginary parts of the eigenvalues of TM as a function of $\omega$. Here $\lambda$ represent eigenvalues of TM. Every eigenvalue is plotted with a different color, and corresponding colors of real and imaginary parts are the same. The vertical dashed lines correspond to the extrema of the energy dispersion. Panel (vi) shows the DOS in the vicinity of the extrema of the band dispersion. The symbols are numerically obtained DOS, while the continuous lines are plot of the analytical result. All energies are in units of $t_1$.}
\end{figure}

\section{Connecting order of EP to exponent of divergence of density of states}
\label{app2}

In this section we provide a detailed derivation that connects the order of EP of transfer matrix with the exponent of divergence of density of states (DOS) near the van Hove singularities. The van Hove singularities  are energies where the density of states diverge. For a single-band system that we are considering in this work, the DOS is given by
\begin{align}
\label{DOS}
\nu(\mu)=\int_{-\pi}^\pi \frac{dk}{2 \pi} \, \delta\big(\mu-\epsilon(k)\big), 
\end{align}
where $\delta(x)$ is the Dirac delta function. 
We can re-write $\nu(\mu)$ as
\begin{align}
\label{DOS2}
    \nu(\mu)=\sum_{r=0}^{q-1}\int_{k_r-\delta k}^{k_r+\delta k} \frac{dk}{2\pi} \, \delta\big(\mu-\epsilon(k)\big),
\end{align}
where $k_r$ are the real solutions to $\mu-\epsilon(k)=0$, given by $\{k_0,k_1,k_2,\ldots,k_{q-1}\}$ with $-\pi \leq k_r \leq \pi ~~\forall ~ r=0,1,2 \ldots q-1$. Note that $\delta k$, which is a small window about $k_r$, that appears in the limits of the integral of Eq.~\eqref{DOS2}, ensures distinct contribution each coming from a particular $k_r$. 

Now, making a change of variable $\omega=\epsilon(k)$, Eq.~\eqref{DOS2} simplifies to
\begin{align}
\label{DOS3}
    \nu(\mu)=\sum_{r=0}^{q-1}\int_{\mu -\delta \mu}^{\mu+\delta \mu} \frac{d\omega}{2\pi}~~\Big| \frac{dk}{d\omega} \Big|_{k\approx k_r} \delta(\mu-\omega) ,
\end{align}
where $\Big|\frac{dk}{d\omega}\Big|_{k \approx k_r}$ implies calculating  $\Big|\frac{dk}{d\omega}\Big|$ in the vicinity of $k_r$, and $\delta \mu$ is a small window of $\mu$. It is worth noting that so-far we have not assumed any additional restrictions on the list $\{k_0,k_1,k_2,\ldots,k_{q-1}\}$. The goal is now to further simplify Eq.~\eqref{DOS3} by considering the situation of having EPs of transfer matrix. 

Let us now consider the DOS when $\mu$ is close to an energy $\omega_0$, such that there is a $p$th order EP of $\mathbf{T}(\omega)$. This means that $\omega_0=\epsilon(k)$ has $p$ repeated solutions, which, say, is given by $k_0$. Furthermore, for simplicity let us assume that there are only two other non-degenerate solutions $k_*$ and $-k_*$ with $0 \leq k_* \leq \pi$. For $k=\pm k_*$,  $\Big|\frac{dk}{d\omega}\Big|$ near $k=k*$ is simply $\Big|\frac{1}{a_1(k_*)}\Big|$ where recall that $a_1(k)= \frac{d\epsilon(k)}{dk}$. To find $\Big|\frac{dk}{d\omega}\Big|$ near $k=k_0$ which is a $p$th order EP, we expand $\epsilon(k)$ around $k_0$ , and using Theorem~\ref{EP_theorem_sup}, obtain, to the leading order in $k-k_0$,
\begin{align}
    \omega\simeq \omega_0+\frac{(k-k_0)^p}{p!} a_p(k_0),
\end{align}
where we have also used $\omega=\epsilon(k)$ and $\omega_0=\epsilon(k_0)$. Solving this equation for $k$ gives
\begin{align}
\label{k_solutions}
    k=
    \begin{cases}
    & k_0 + \Big| \frac{(\omega -\omega_0)p!}{a_p(k_0)} \Big|^{\frac{1}{p}} e^{ \frac{2\pi m i}{p}},~\forall~\frac{(\omega -\omega_0)}{a_p(k_0)}>0 \\
    & k_0 + \Big| \frac{(\omega -\omega_0)p!}{a_p(k_0)} \Big|^{\frac{1}{p}} e^{\frac{(2m+1)\pi i}{p}},~\forall~\frac{(\omega -\omega_0)}{a_p(k_0)}<0
    \end{cases},
\end{align}
with $m=0,1,\ldots, p-1$.  There can be various cases, as discussed in the main text. In the following, we consider the case where $k_0$ corresponds to a band minimum. In this case, $p$ even, i.e, $p=2s$ and $a_{2s}(k_0)>0$. Other cases can be considered similarly. In Eq.~\eqref{DOS3}, only real values of $k$ matters. For $a_{2s}(k_0)>0$, we see from Eq.\eqref{k_solutions} that two real values occur when $\omega-\omega_0>0$, corresponding to $m=0$ and $m=s$, while no there are no real values of $k$ for $\omega-\omega_0<0$. Thus, the DOS is affected by the EP only when it is approached from above. With  $\omega-\omega_0>0$, for both real values of $k$, from Eq.~\eqref{k_solutions}, we find that
\begin{align}
    \Big| \frac{dk}{d\omega} \Big|_{k\approx k_0}=\frac{(p-1)!}{a_p(k_0) }\left[\frac{a_p(k_0)}{p! (\omega -\omega_0)}\right]^{\frac{p-1}{p}}.
\end{align}
Using this in Eq.\eqref{DOS3} for both real values of $k$, we get 
\begin{align}
\nu(\mu) = \frac{(p-1)!}{\pi a_p(k_0) }\left[\frac{a_p(k_0)}{p! (\mu -\omega_0)}\right]^{\frac{p-1}{p}} + \frac{1}{\pi a_1(k_*)},
\end{align}
where we have also added the contributions from $k=\pm k_*$. This connects the exponent of divergence of DOS with the order of EP of TM.

\section{Numerical procedure for computing density of states}
\label{sec:num}
In this section, we provide the numerical procedure to compute DOS. To numerically obtain the DOS, it is easier to first obtain the integrated DOS, defined as
\begin{align}
    D(E)=\int_{-\pi}^\pi \frac{dk}{2 \pi} \, \Theta\big(E-\epsilon(k)\big),
\end{align}
where $\Theta(x)$ is the Heaviside theta function. The DOS is then obtained by 
\begin{align}
    \nu(\mu)=\frac{d D(E)}{dE}\Big|_{E=\mu}.
    \label{DOS-numer}
\end{align}
We make a very fine grid of $k$ between $-\pi$ to $\pi$, in steps of $\delta k$, and sample $\epsilon(k)$ at these points. The $D(E)$ is then simply given by the number of $k$-points satisfying $\epsilon(k)<E$, multiplied by $\delta k$ and divided by $2\pi$. After obtaining $D(E)$, $\nu(\mu)$ in Eq.~\eqref{DOS-numer} is obtained by calculating the derivative using finite difference method. To get highly accurate derivatives, we used fourth order central finite difference method. 
\\

\section{Results for lattice range of hopping $n=3$ for sixth order EP of transfer matrix}
\label{sec:n3}
In this section we provide numerical results in presence of the highest order exceptional point (EP) for range of hopping $n=3$. For $n=3$, the allowed orders of EP of the TM are second, third, fourth and sixth. Fifth order EP is not possible for $n=3$ (see Theorem 2 of the main text). In the main text, we have given plots for cases with third and fourth order EPs of TM. Here we give the plots for a case with sixth order (highest order) EP.

As discussed in the main text, for $n=3$, a sixth order EP of TM occurs at $k=\pi$, when the hoppings satisfy $t_2=2t_1/5, t_3=t_1/15$. In Fig.~\ref{fig:sixth_order}(i), we show the plot of the dispersion relation $\epsilon(k)$ for these parameters. We see that there are only two critical points, corresponding to the extrema of band dispersion, occurring at $k=0$, and $k=\pi$. In Fig.~\ref{fig:sixth_order}(ii), we show the plot of DOS $\nu(\omega)$. We clearly see that $\nu(\omega)$ diverges while approaching the minimum of the band from above and the maximum of the band from below. In Fig.~\ref{fig:sixth_order}(iii), we plot the number of eigenvalues of TM that are of unit modulus, as a function of $\omega$. We see that this quantity increases on crossing a minimum from below, while it decreases on crossing a maximum from below. In Fig.~\ref{fig:sixth_order}(iv) and (v), we plot the real and the imaginary parts eigenvalues of TM. We see that at the band minimum, occurring at $k=0$, both the real and the imaginary parts of two eigenvalues coalesce. This shows that it is an EP of second order. At the band maximum, occurring at $k=\pi$, both the real and the imaginary parts of all the six eigenvalues of TM coalesce. This shows it is an EP of sixth order. Finally, in Fig.~\ref{fig:sixth_order}(vi), we show the plot of DOS in the vicinity of the band minimum and maximum. We clearly see that DOS diverges as $\delta^{1/2}$ at band minimum (the second order EP), and as $\delta^{5/6}$ at band maximum (the sixth order EP).

\bibliography{ref_isolated}

\begin{thebibliography}{58}%
\makeatletter
\providecommand \@ifxundefined [1]{%
 \@ifx{#1\undefined}
}%
\providecommand \@ifnum [1]{%
 \ifnum #1\expandafter \@firstoftwo
 \else \expandafter \@secondoftwo
 \fi
}%
\providecommand \@ifx [1]{%
 \ifx #1\expandafter \@firstoftwo
 \else \expandafter \@secondoftwo
 \fi
}%
\providecommand \natexlab [1]{#1}%
\providecommand \enquote  [1]{``#1''}%
\providecommand \bibnamefont  [1]{#1}%
\providecommand \bibfnamefont [1]{#1}%
\providecommand \citenamefont [1]{#1}%
\providecommand \href@noop [0]{\@secondoftwo}%
\providecommand \href [0]{\begingroup \@sanitize@url \@href}%
\providecommand \@href[1]{\@@startlink{#1}\@@href}%
\providecommand \@@href[1]{\endgroup#1\@@endlink}%
\providecommand \@sanitize@url [0]{\catcode `\\12\catcode `\$12\catcode
  `\&12\catcode `\#12\catcode `\^12\catcode `\_12\catcode `\%12\relax}%
\providecommand \@@startlink[1]{}%
\providecommand \@@endlink[0]{}%
\providecommand \url  [0]{\begingroup\@sanitize@url \@url }%
\providecommand \@url [1]{\endgroup\@href {#1}{\urlprefix }}%
\providecommand \urlprefix  [0]{URL }%
\providecommand \Eprint [0]{\href }%
\providecommand \doibase [0]{https://doi.org/}%
\providecommand \selectlanguage [0]{\@gobble}%
\providecommand \bibinfo  [0]{\@secondoftwo}%
\providecommand \bibfield  [0]{\@secondoftwo}%
\providecommand \translation [1]{[#1]}%
\providecommand \BibitemOpen [0]{}%
\providecommand \bibitemStop [0]{}%
\providecommand \bibitemNoStop [0]{.\EOS\space}%
\providecommand \EOS [0]{\spacefactor3000\relax}%
\providecommand \BibitemShut  [1]{\csname bibitem#1\endcsname}%
\let\auto@bib@innerbib\@empty
\bibitem [{\citenamefont {Nishimori}\ and\ \citenamefont
  {Ortiz}(2010)}]{singularity_book}%
  \BibitemOpen
  \bibfield  {author} {\bibinfo {author} {\bibfnamefont {H.}~\bibnamefont
  {Nishimori}}\ and\ \bibinfo {author} {\bibfnamefont {G.}~\bibnamefont
  {Ortiz}},\ }\href {https://doi.org/10.1093/acprof:oso/9780199577224.001.0001}
  {\emph {\bibinfo {title} {{Elements of Phase Transitions and Critical
  Phenomena}}}}\ (\bibinfo  {publisher} {Oxford University Press},\ \bibinfo
  {year} {2010})\BibitemShut {NoStop}%
\bibitem [{\citenamefont {Sachdev}(2011)}]{Sachdev_2011}%
  \BibitemOpen
  \bibfield  {author} {\bibinfo {author} {\bibfnamefont {S.}~\bibnamefont
  {Sachdev}},\ }\href@noop {} {\emph {\bibinfo {title} {Quantum Phase
  Transitions}}},\ \bibinfo {edition} {2nd}\ ed.\ (\bibinfo  {publisher}
  {Cambridge University Press},\ \bibinfo {year} {2011})\BibitemShut {NoStop}%
\bibitem [{\citenamefont {Yeomans}(1992)}]{yeomans1992statistical}%
  \BibitemOpen
  \bibfield  {author} {\bibinfo {author} {\bibfnamefont {J.~M.}\ \bibnamefont
  {Yeomans}},\ }\href@noop {} {\emph {\bibinfo {title} {Statistical mechanics
  of phase transitions}}}\ (\bibinfo  {publisher} {Clarendon Press},\ \bibinfo
  {year} {1992})\BibitemShut {NoStop}%
\bibitem [{\citenamefont {Van~Hove}(1953)}]{van_hove_1953}%
  \BibitemOpen
  \bibfield  {author} {\bibinfo {author} {\bibfnamefont {L.}~\bibnamefont
  {Van~Hove}},\ }\bibfield  {title} {\bibinfo {title} {The occurrence of
  singularities in the elastic frequency distribution of a crystal},\ }\href
  {https://doi.org/10.1103/PhysRev.89.1189} {\bibfield  {journal} {\bibinfo
  {journal} {Phys. Rev.}\ }\textbf {\bibinfo {volume} {89}},\ \bibinfo {pages}
  {1189} (\bibinfo {year} {1953})}\BibitemShut {NoStop}%
\bibitem [{\citenamefont {Ashcroft}\ and\ \citenamefont
  {Mermin}(1976)}]{Ashcroft}%
  \BibitemOpen
  \bibfield  {author} {\bibinfo {author} {\bibfnamefont {N.~W.}\ \bibnamefont
  {Ashcroft}}\ and\ \bibinfo {author} {\bibfnamefont {D.~N.}\ \bibnamefont
  {Mermin}},\ }\href@noop {} {\emph {\bibinfo {title} {Solid State Physics}}}\
  (\bibinfo  {publisher} {New York: Holt, Rinehart and Winston},\ \bibinfo
  {year} {1976})\BibitemShut {NoStop}%
\bibitem [{\citenamefont {Mahan}(2000)}]{Mahan}%
  \BibitemOpen
  \bibfield  {author} {\bibinfo {author} {\bibfnamefont {G.~D.}\ \bibnamefont
  {Mahan}},\ }\href@noop {} {\emph {\bibinfo {title} {Many-Particle Physics}}}\
  (\bibinfo  {publisher} {Springer New York, NY},\ \bibinfo {year}
  {2000})\BibitemShut {NoStop}%
\bibitem [{\citenamefont {Shtyk}\ \emph {et~al.}(2017)\citenamefont {Shtyk},
  \citenamefont {Goldstein},\ and\ \citenamefont {Chamon}}]{Shtyk_2017}%
  \BibitemOpen
  \bibfield  {author} {\bibinfo {author} {\bibfnamefont {A.}~\bibnamefont
  {Shtyk}}, \bibinfo {author} {\bibfnamefont {G.}~\bibnamefont {Goldstein}},\
  and\ \bibinfo {author} {\bibfnamefont {C.}~\bibnamefont {Chamon}},\
  }\bibfield  {title} {\bibinfo {title} {Electrons at the monkey saddle: A
  multicritical lifshitz point},\ }\href
  {https://doi.org/10.1103/PhysRevB.95.035137} {\bibfield  {journal} {\bibinfo
  {journal} {Phys. Rev. B}\ }\textbf {\bibinfo {volume} {95}},\ \bibinfo
  {pages} {035137} (\bibinfo {year} {2017})}\BibitemShut {NoStop}%
\bibitem [{\citenamefont {Zhang}\ \emph {et~al.}(2017)\citenamefont {Zhang},
  \citenamefont {Liu}, \citenamefont {Nshimiyimana}, \citenamefont {Deng},
  \citenamefont {Hu}, \citenamefont {Chi}, \citenamefont {Wu}, \citenamefont
  {Liu}, \citenamefont {Chu},\ and\ \citenamefont {Sun}}]{1D_van_hove}%
  \BibitemOpen
  \bibfield  {author} {\bibinfo {author} {\bibfnamefont {J.}~\bibnamefont
  {Zhang}}, \bibinfo {author} {\bibfnamefont {S.}~\bibnamefont {Liu}}, \bibinfo
  {author} {\bibfnamefont {J.~P.}\ \bibnamefont {Nshimiyimana}}, \bibinfo
  {author} {\bibfnamefont {Y.}~\bibnamefont {Deng}}, \bibinfo {author}
  {\bibfnamefont {X.}~\bibnamefont {Hu}}, \bibinfo {author} {\bibfnamefont
  {X.}~\bibnamefont {Chi}}, \bibinfo {author} {\bibfnamefont {P.}~\bibnamefont
  {Wu}}, \bibinfo {author} {\bibfnamefont {J.}~\bibnamefont {Liu}}, \bibinfo
  {author} {\bibfnamefont {W.}~\bibnamefont {Chu}},\ and\ \bibinfo {author}
  {\bibfnamefont {L.}~\bibnamefont {Sun}},\ }\bibfield  {title} {\bibinfo
  {title} {Observation of van hove singularities and temperature dependence of
  electrical characteristics in suspended carbon nanotube schottky barrier
  transistors},\ }\href
  {https://doi.org/https://doi.org/10.1007/s40820-017-0171-3} {\bibfield
  {journal} {\bibinfo  {journal} {Nano-Micro Letters}\ }\textbf {\bibinfo
  {volume} {10}},\ \bibinfo {pages} {25} (\bibinfo {year} {2017})}\BibitemShut
  {NoStop}%
\bibitem [{\citenamefont {Yuan}\ \emph {et~al.}(2019)\citenamefont {Yuan},
  \citenamefont {Isobe},\ and\ \citenamefont {Fu}}]{Yuan_2019}%
  \BibitemOpen
  \bibfield  {author} {\bibinfo {author} {\bibfnamefont {N.~F.~Q.}\
  \bibnamefont {Yuan}}, \bibinfo {author} {\bibfnamefont {H.}~\bibnamefont
  {Isobe}},\ and\ \bibinfo {author} {\bibfnamefont {L.}~\bibnamefont {Fu}},\
  }\bibfield  {title} {\bibinfo {title} {Magic of high-order van hove
  singularity},\ }\href {https://doi.org/10.1038/s41467-019-13670-9} {\bibfield
   {journal} {\bibinfo  {journal} {Nature Communications}\ }\textbf {\bibinfo
  {volume} {10}},\ \bibinfo {pages} {5769} (\bibinfo {year}
  {2019})}\BibitemShut {NoStop}%
\bibitem [{\citenamefont {Scammell}\ \emph {et~al.}(2023)\citenamefont
  {Scammell}, \citenamefont {Ingham}, \citenamefont {Li},\ and\ \citenamefont
  {Sushkov}}]{Scammell_2D_van_hove}%
  \BibitemOpen
  \bibfield  {author} {\bibinfo {author} {\bibfnamefont {H.~D.}\ \bibnamefont
  {Scammell}}, \bibinfo {author} {\bibfnamefont {J.}~\bibnamefont {Ingham}},
  \bibinfo {author} {\bibfnamefont {T.}~\bibnamefont {Li}},\ and\ \bibinfo
  {author} {\bibfnamefont {O.~P.}\ \bibnamefont {Sushkov}},\ }\bibfield
  {title} {\bibinfo {title} {Chiral excitonic order from twofold van hove
  singularities in kagome metals},\ }\href
  {https://doi.org/https://doi.org/10.1038/s41467-023-35987-2} {\bibfield
  {journal} {\bibinfo  {journal} {Nature Communications}\ }\textbf {\bibinfo
  {volume} {14}},\ \bibinfo {pages} {605} (\bibinfo {year} {2023})}\BibitemShut
  {NoStop}%
\bibitem [{\citenamefont {Yuan}\ and\ \citenamefont
  {Fu}(2020)}]{critical_topology_Liang}%
  \BibitemOpen
  \bibfield  {author} {\bibinfo {author} {\bibfnamefont {N.~F.~Q.}\
  \bibnamefont {Yuan}}\ and\ \bibinfo {author} {\bibfnamefont {L.}~\bibnamefont
  {Fu}},\ }\bibfield  {title} {\bibinfo {title} {Classification of critical
  points in energy bands based on topology, scaling, and symmetry},\ }\href
  {https://doi.org/10.1103/PhysRevB.101.125120} {\bibfield  {journal} {\bibinfo
   {journal} {Phys. Rev. B}\ }\textbf {\bibinfo {volume} {101}},\ \bibinfo
  {pages} {125120} (\bibinfo {year} {2020})}\BibitemShut {NoStop}%
\bibitem [{\citenamefont {Efremov}\ \emph {et~al.}(2019)\citenamefont
  {Efremov}, \citenamefont {Shtyk}, \citenamefont {Rost}, \citenamefont
  {Chamon}, \citenamefont {Mackenzie},\ and\ \citenamefont
  {Betouras}}]{Efremov_2019}%
  \BibitemOpen
  \bibfield  {author} {\bibinfo {author} {\bibfnamefont {D.~V.}\ \bibnamefont
  {Efremov}}, \bibinfo {author} {\bibfnamefont {A.}~\bibnamefont {Shtyk}},
  \bibinfo {author} {\bibfnamefont {A.~W.}\ \bibnamefont {Rost}}, \bibinfo
  {author} {\bibfnamefont {C.}~\bibnamefont {Chamon}}, \bibinfo {author}
  {\bibfnamefont {A.~P.}\ \bibnamefont {Mackenzie}},\ and\ \bibinfo {author}
  {\bibfnamefont {J.~J.}\ \bibnamefont {Betouras}},\ }\bibfield  {title}
  {\bibinfo {title} {Multicritical fermi surface topological transitions},\
  }\href {https://doi.org/10.1103/PhysRevLett.123.207202} {\bibfield  {journal}
  {\bibinfo  {journal} {Phys. Rev. Lett.}\ }\textbf {\bibinfo {volume} {123}},\
  \bibinfo {pages} {207202} (\bibinfo {year} {2019})}\BibitemShut {NoStop}%
\bibitem [{\citenamefont {Classen}\ \emph {et~al.}(2020)\citenamefont
  {Classen}, \citenamefont {Chubukov}, \citenamefont {Honerkamp},\ and\
  \citenamefont {Scherer}}]{Classen_2020}%
  \BibitemOpen
  \bibfield  {author} {\bibinfo {author} {\bibfnamefont {L.}~\bibnamefont
  {Classen}}, \bibinfo {author} {\bibfnamefont {A.~V.}\ \bibnamefont
  {Chubukov}}, \bibinfo {author} {\bibfnamefont {C.}~\bibnamefont
  {Honerkamp}},\ and\ \bibinfo {author} {\bibfnamefont {M.~M.}\ \bibnamefont
  {Scherer}},\ }\bibfield  {title} {\bibinfo {title} {Competing orders at
  higher-order van hove points},\ }\href
  {https://doi.org/10.1103/PhysRevB.102.125141} {\bibfield  {journal} {\bibinfo
   {journal} {Phys. Rev. B}\ }\textbf {\bibinfo {volume} {102}},\ \bibinfo
  {pages} {125141} (\bibinfo {year} {2020})}\BibitemShut {NoStop}%
\bibitem [{\citenamefont {Salamon}\ \emph {et~al.}(2020)\citenamefont
  {Salamon}, \citenamefont {Celi}, \citenamefont {Chhajlany}, \citenamefont
  {Fr\'erot}, \citenamefont {Lewenstein}, \citenamefont {Tarruell},\ and\
  \citenamefont {Rakshit}}]{Salamon_2020}%
  \BibitemOpen
  \bibfield  {author} {\bibinfo {author} {\bibfnamefont {T.}~\bibnamefont
  {Salamon}}, \bibinfo {author} {\bibfnamefont {A.}~\bibnamefont {Celi}},
  \bibinfo {author} {\bibfnamefont {R.~W.}\ \bibnamefont {Chhajlany}}, \bibinfo
  {author} {\bibfnamefont {I.}~\bibnamefont {Fr\'erot}}, \bibinfo {author}
  {\bibfnamefont {M.}~\bibnamefont {Lewenstein}}, \bibinfo {author}
  {\bibfnamefont {L.}~\bibnamefont {Tarruell}},\ and\ \bibinfo {author}
  {\bibfnamefont {D.}~\bibnamefont {Rakshit}},\ }\bibfield  {title} {\bibinfo
  {title} {Simulating twistronics without a twist},\ }\href
  {https://doi.org/10.1103/PhysRevLett.125.030504} {\bibfield  {journal}
  {\bibinfo  {journal} {Phys. Rev. Lett.}\ }\textbf {\bibinfo {volume} {125}},\
  \bibinfo {pages} {030504} (\bibinfo {year} {2020})}\BibitemShut {NoStop}%
\bibitem [{\citenamefont {Chandrasekaran}\ \emph {et~al.}(2020)\citenamefont
  {Chandrasekaran}, \citenamefont {Shtyk}, \citenamefont {Betouras},\ and\
  \citenamefont {Chamon}}]{Chadrasekaran_2020}%
  \BibitemOpen
  \bibfield  {author} {\bibinfo {author} {\bibfnamefont {A.}~\bibnamefont
  {Chandrasekaran}}, \bibinfo {author} {\bibfnamefont {A.}~\bibnamefont
  {Shtyk}}, \bibinfo {author} {\bibfnamefont {J.~J.}\ \bibnamefont
  {Betouras}},\ and\ \bibinfo {author} {\bibfnamefont {C.}~\bibnamefont
  {Chamon}},\ }\bibfield  {title} {\bibinfo {title} {Catastrophe theory
  classification of fermi surface topological transitions in two dimensions},\
  }\href {https://doi.org/10.1103/PhysRevResearch.2.013355} {\bibfield
  {journal} {\bibinfo  {journal} {Phys. Rev. Res.}\ }\textbf {\bibinfo {volume}
  {2}},\ \bibinfo {pages} {013355} (\bibinfo {year} {2020})}\BibitemShut
  {NoStop}%
\bibitem [{\citenamefont {Guerci}\ \emph {et~al.}(2022)\citenamefont {Guerci},
  \citenamefont {Simon},\ and\ \citenamefont {Mora}}]{Guerci_2022}%
  \BibitemOpen
  \bibfield  {author} {\bibinfo {author} {\bibfnamefont {D.}~\bibnamefont
  {Guerci}}, \bibinfo {author} {\bibfnamefont {P.}~\bibnamefont {Simon}},\ and\
  \bibinfo {author} {\bibfnamefont {C.}~\bibnamefont {Mora}},\ }\bibfield
  {title} {\bibinfo {title} {Higher-order van hove singularity in magic-angle
  twisted trilayer graphene},\ }\href
  {https://doi.org/10.1103/PhysRevResearch.4.L012013} {\bibfield  {journal}
  {\bibinfo  {journal} {Phys. Rev. Res.}\ }\textbf {\bibinfo {volume} {4}},\
  \bibinfo {pages} {L012013} (\bibinfo {year} {2022})}\BibitemShut {NoStop}%
\bibitem [{\citenamefont {Hu}\ \emph {et~al.}(2022)\citenamefont {Hu},
  \citenamefont {Wu}, \citenamefont {Ortiz}, \citenamefont {Ju}, \citenamefont
  {Han}, \citenamefont {Ma}, \citenamefont {Plumb}, \citenamefont {Radovic},
  \citenamefont {Thomale}, \citenamefont {Wilson}, \citenamefont {Schnyder},\
  and\ \citenamefont {Shi}}]{Hu_2022}%
  \BibitemOpen
  \bibfield  {author} {\bibinfo {author} {\bibfnamefont {Y.}~\bibnamefont
  {Hu}}, \bibinfo {author} {\bibfnamefont {X.}~\bibnamefont {Wu}}, \bibinfo
  {author} {\bibfnamefont {B.~R.}\ \bibnamefont {Ortiz}}, \bibinfo {author}
  {\bibfnamefont {S.}~\bibnamefont {Ju}}, \bibinfo {author} {\bibfnamefont
  {X.}~\bibnamefont {Han}}, \bibinfo {author} {\bibfnamefont {J.}~\bibnamefont
  {Ma}}, \bibinfo {author} {\bibfnamefont {N.~C.}\ \bibnamefont {Plumb}},
  \bibinfo {author} {\bibfnamefont {M.}~\bibnamefont {Radovic}}, \bibinfo
  {author} {\bibfnamefont {R.}~\bibnamefont {Thomale}}, \bibinfo {author}
  {\bibfnamefont {S.~D.}\ \bibnamefont {Wilson}}, \bibinfo {author}
  {\bibfnamefont {A.~P.}\ \bibnamefont {Schnyder}},\ and\ \bibinfo {author}
  {\bibfnamefont {M.}~\bibnamefont {Shi}},\ }\bibfield  {title} {\bibinfo
  {title} {Rich nature of van hove singularities in kagome superconductor
  csv3sb5},\ }\href {https://doi.org/10.1038/s41467-022-29828-x} {\bibfield
  {journal} {\bibinfo  {journal} {Nature Communications}\ }\textbf {\bibinfo
  {volume} {13}},\ \bibinfo {pages} {2220} (\bibinfo {year}
  {2022})}\BibitemShut {NoStop}%
\bibitem [{\citenamefont {Chandrasekaran}\ and\ \citenamefont
  {Betouras}(2022)}]{Chandrasekaran_2022}%
  \BibitemOpen
  \bibfield  {author} {\bibinfo {author} {\bibfnamefont {A.}~\bibnamefont
  {Chandrasekaran}}\ and\ \bibinfo {author} {\bibfnamefont {J.~J.}\
  \bibnamefont {Betouras}},\ }\bibfield  {title} {\bibinfo {title} {Effect of
  disorder on density of states and conductivity in higher-order van hove
  singularities in two-dimensional bands},\ }\href
  {https://doi.org/10.1103/PhysRevB.105.075144} {\bibfield  {journal} {\bibinfo
   {journal} {Phys. Rev. B}\ }\textbf {\bibinfo {volume} {105}},\ \bibinfo
  {pages} {075144} (\bibinfo {year} {2022})}\BibitemShut {NoStop}%
\bibitem [{\citenamefont {Shankar}\ \emph {et~al.}(2023)\citenamefont
  {Shankar}, \citenamefont {Oriekhov}, \citenamefont {Mitchell},\ and\
  \citenamefont {Fritz}}]{Shankar_2023}%
  \BibitemOpen
  \bibfield  {author} {\bibinfo {author} {\bibfnamefont {A.~S.}\ \bibnamefont
  {Shankar}}, \bibinfo {author} {\bibfnamefont {D.~O.}\ \bibnamefont
  {Oriekhov}}, \bibinfo {author} {\bibfnamefont {A.~K.}\ \bibnamefont
  {Mitchell}},\ and\ \bibinfo {author} {\bibfnamefont {L.}~\bibnamefont
  {Fritz}},\ }\bibfield  {title} {\bibinfo {title} {Kondo effect in twisted
  bilayer graphene},\ }\href {https://doi.org/10.1103/PhysRevB.107.245102}
  {\bibfield  {journal} {\bibinfo  {journal} {Phys. Rev. B}\ }\textbf {\bibinfo
  {volume} {107}},\ \bibinfo {pages} {245102} (\bibinfo {year}
  {2023})}\BibitemShut {NoStop}%
\bibitem [{\citenamefont {Sheffer}\ \emph {et~al.}(2023)\citenamefont
  {Sheffer}, \citenamefont {Queiroz},\ and\ \citenamefont
  {Stern}}]{Sheffer_2023}%
  \BibitemOpen
  \bibfield  {author} {\bibinfo {author} {\bibfnamefont {Y.}~\bibnamefont
  {Sheffer}}, \bibinfo {author} {\bibfnamefont {R.}~\bibnamefont {Queiroz}},\
  and\ \bibinfo {author} {\bibfnamefont {A.}~\bibnamefont {Stern}},\ }\bibfield
   {title} {\bibinfo {title} {Symmetries as the guiding principle for
  flattening bands of dirac fermions},\ }\href
  {https://doi.org/10.1103/PhysRevX.13.021012} {\bibfield  {journal} {\bibinfo
  {journal} {Phys. Rev. X}\ }\textbf {\bibinfo {volume} {13}},\ \bibinfo
  {pages} {021012} (\bibinfo {year} {2023})}\BibitemShut {NoStop}%
\bibitem [{\citenamefont {Zakharov}\ \emph {et~al.}(2024)\citenamefont
  {Zakharov}, \citenamefont {Bozkurt}, \citenamefont {Akhmerov},\ and\
  \citenamefont {Oriekhov}}]{Zakharov_2024}%
  \BibitemOpen
  \bibfield  {author} {\bibinfo {author} {\bibfnamefont {V.~A.}\ \bibnamefont
  {Zakharov}}, \bibinfo {author} {\bibfnamefont {A.~M.}\ \bibnamefont
  {Bozkurt}}, \bibinfo {author} {\bibfnamefont {A.~R.}\ \bibnamefont
  {Akhmerov}},\ and\ \bibinfo {author} {\bibfnamefont {D.~O.}\ \bibnamefont
  {Oriekhov}},\ }\bibfield  {title} {\bibinfo {title} {Landau quantization near
  generalized van hove singularities: Magnetic breakdown and orbit networks},\
  }\href {https://doi.org/10.1103/PhysRevB.109.L081103} {\bibfield  {journal}
  {\bibinfo  {journal} {Phys. Rev. B}\ }\textbf {\bibinfo {volume} {109}},\
  \bibinfo {pages} {L081103} (\bibinfo {year} {2024})}\BibitemShut {NoStop}%
\bibitem [{\citenamefont {Chen}\ \emph {et~al.}(2022)\citenamefont {Chen},
  \citenamefont {Abbasi}, \citenamefont {Ha}, \citenamefont {Erdamar},
  \citenamefont {Joglekar},\ and\ \citenamefont {Murch}}]{Chen_2022}%
  \BibitemOpen
  \bibfield  {author} {\bibinfo {author} {\bibfnamefont {W.}~\bibnamefont
  {Chen}}, \bibinfo {author} {\bibfnamefont {M.}~\bibnamefont {Abbasi}},
  \bibinfo {author} {\bibfnamefont {B.}~\bibnamefont {Ha}}, \bibinfo {author}
  {\bibfnamefont {S.}~\bibnamefont {Erdamar}}, \bibinfo {author} {\bibfnamefont
  {Y.~N.}\ \bibnamefont {Joglekar}},\ and\ \bibinfo {author} {\bibfnamefont
  {K.~W.}\ \bibnamefont {Murch}},\ }\bibfield  {title} {\bibinfo {title}
  {Decoherence-induced exceptional points in a dissipative superconducting
  qubit},\ }\href {https://doi.org/10.1103/PhysRevLett.128.110402} {\bibfield
  {journal} {\bibinfo  {journal} {Phys. Rev. Lett.}\ }\textbf {\bibinfo
  {volume} {128}},\ \bibinfo {pages} {110402} (\bibinfo {year}
  {2022})}\BibitemShut {NoStop}%
\bibitem [{\citenamefont {Abbasi}\ \emph {et~al.}(2022)\citenamefont {Abbasi},
  \citenamefont {Chen}, \citenamefont {Naghiloo}, \citenamefont {Joglekar},\
  and\ \citenamefont {Murch}}]{Abbasi_2022}%
  \BibitemOpen
  \bibfield  {author} {\bibinfo {author} {\bibfnamefont {M.}~\bibnamefont
  {Abbasi}}, \bibinfo {author} {\bibfnamefont {W.}~\bibnamefont {Chen}},
  \bibinfo {author} {\bibfnamefont {M.}~\bibnamefont {Naghiloo}}, \bibinfo
  {author} {\bibfnamefont {Y.~N.}\ \bibnamefont {Joglekar}},\ and\ \bibinfo
  {author} {\bibfnamefont {K.~W.}\ \bibnamefont {Murch}},\ }\bibfield  {title}
  {\bibinfo {title} {Topological quantum state control through
  exceptional-point proximity},\ }\href
  {https://doi.org/10.1103/PhysRevLett.128.160401} {\bibfield  {journal}
  {\bibinfo  {journal} {Phys. Rev. Lett.}\ }\textbf {\bibinfo {volume} {128}},\
  \bibinfo {pages} {160401} (\bibinfo {year} {2022})}\BibitemShut {NoStop}%
\bibitem [{\citenamefont {Liao}\ \emph {et~al.}(2021)\citenamefont {Liao},
  \citenamefont {Leblanc}, \citenamefont {Ren}, \citenamefont {Li},
  \citenamefont {Li}, \citenamefont {Solnyshkov}, \citenamefont {Malpuech},
  \citenamefont {Yao},\ and\ \citenamefont {Fu}}]{Liao_2021}%
  \BibitemOpen
  \bibfield  {author} {\bibinfo {author} {\bibfnamefont {Q.}~\bibnamefont
  {Liao}}, \bibinfo {author} {\bibfnamefont {C.}~\bibnamefont {Leblanc}},
  \bibinfo {author} {\bibfnamefont {J.}~\bibnamefont {Ren}}, \bibinfo {author}
  {\bibfnamefont {F.}~\bibnamefont {Li}}, \bibinfo {author} {\bibfnamefont
  {Y.}~\bibnamefont {Li}}, \bibinfo {author} {\bibfnamefont {D.}~\bibnamefont
  {Solnyshkov}}, \bibinfo {author} {\bibfnamefont {G.}~\bibnamefont
  {Malpuech}}, \bibinfo {author} {\bibfnamefont {J.}~\bibnamefont {Yao}},\ and\
  \bibinfo {author} {\bibfnamefont {H.}~\bibnamefont {Fu}},\ }\bibfield
  {title} {\bibinfo {title} {Experimental measurement of the divergent quantum
  metric of an exceptional point},\ }\href
  {https://doi.org/10.1103/PhysRevLett.127.107402} {\bibfield  {journal}
  {\bibinfo  {journal} {Phys. Rev. Lett.}\ }\textbf {\bibinfo {volume} {127}},\
  \bibinfo {pages} {107402} (\bibinfo {year} {2021})}\BibitemShut {NoStop}%
\bibitem [{\citenamefont {Liu}\ \emph {et~al.}(2021)\citenamefont {Liu},
  \citenamefont {Wu}, \citenamefont {Duan}, \citenamefont {Rong},\ and\
  \citenamefont {Du}}]{Liu_2021}%
  \BibitemOpen
  \bibfield  {author} {\bibinfo {author} {\bibfnamefont {W.}~\bibnamefont
  {Liu}}, \bibinfo {author} {\bibfnamefont {Y.}~\bibnamefont {Wu}}, \bibinfo
  {author} {\bibfnamefont {C.-K.}\ \bibnamefont {Duan}}, \bibinfo {author}
  {\bibfnamefont {X.}~\bibnamefont {Rong}},\ and\ \bibinfo {author}
  {\bibfnamefont {J.}~\bibnamefont {Du}},\ }\bibfield  {title} {\bibinfo
  {title} {Dynamically encircling an exceptional point in a real quantum
  system},\ }\href {https://doi.org/10.1103/PhysRevLett.126.170506} {\bibfield
  {journal} {\bibinfo  {journal} {Phys. Rev. Lett.}\ }\textbf {\bibinfo
  {volume} {126}},\ \bibinfo {pages} {170506} (\bibinfo {year}
  {2021})}\BibitemShut {NoStop}%
\bibitem [{\citenamefont {Chen}\ \emph {et~al.}(2020)\citenamefont {Chen},
  \citenamefont {Liu}, \citenamefont {Luan}, \citenamefont {Liu}, \citenamefont
  {Wang}, \citenamefont {Zhu}, \citenamefont {Li}, \citenamefont {Gu},
  \citenamefont {Liang}, \citenamefont {Gao}, \citenamefont {Lu}, \citenamefont
  {Ge}, \citenamefont {Zhang}, \citenamefont {Zhu},\ and\ \citenamefont
  {Ma}}]{Chen_2020}%
  \BibitemOpen
  \bibfield  {author} {\bibinfo {author} {\bibfnamefont {H.-Z.}\ \bibnamefont
  {Chen}}, \bibinfo {author} {\bibfnamefont {T.}~\bibnamefont {Liu}}, \bibinfo
  {author} {\bibfnamefont {H.-Y.}\ \bibnamefont {Luan}}, \bibinfo {author}
  {\bibfnamefont {R.-J.}\ \bibnamefont {Liu}}, \bibinfo {author} {\bibfnamefont
  {X.-Y.}\ \bibnamefont {Wang}}, \bibinfo {author} {\bibfnamefont {X.-F.}\
  \bibnamefont {Zhu}}, \bibinfo {author} {\bibfnamefont {Y.-B.}\ \bibnamefont
  {Li}}, \bibinfo {author} {\bibfnamefont {Z.-M.}\ \bibnamefont {Gu}}, \bibinfo
  {author} {\bibfnamefont {S.-J.}\ \bibnamefont {Liang}}, \bibinfo {author}
  {\bibfnamefont {H.}~\bibnamefont {Gao}}, \bibinfo {author} {\bibfnamefont
  {L.}~\bibnamefont {Lu}}, \bibinfo {author} {\bibfnamefont {L.}~\bibnamefont
  {Ge}}, \bibinfo {author} {\bibfnamefont {S.}~\bibnamefont {Zhang}}, \bibinfo
  {author} {\bibfnamefont {J.}~\bibnamefont {Zhu}},\ and\ \bibinfo {author}
  {\bibfnamefont {R.-M.}\ \bibnamefont {Ma}},\ }\bibfield  {title} {\bibinfo
  {title} {Revealing the missing dimension at an exceptional point},\ }\href
  {https://doi.org/10.1038/s41567-020-0807-y} {\bibfield  {journal} {\bibinfo
  {journal} {Nature Physics}\ }\textbf {\bibinfo {volume} {16}},\ \bibinfo
  {pages} {571} (\bibinfo {year} {2020})}\BibitemShut {NoStop}%
\bibitem [{\citenamefont {Naghiloo}\ \emph {et~al.}(2019)\citenamefont
  {Naghiloo}, \citenamefont {Abbasi}, \citenamefont {Joglekar},\ and\
  \citenamefont {Murch}}]{Naghiloo_2019}%
  \BibitemOpen
  \bibfield  {author} {\bibinfo {author} {\bibfnamefont {M.}~\bibnamefont
  {Naghiloo}}, \bibinfo {author} {\bibfnamefont {M.}~\bibnamefont {Abbasi}},
  \bibinfo {author} {\bibfnamefont {Y.~N.}\ \bibnamefont {Joglekar}},\ and\
  \bibinfo {author} {\bibfnamefont {K.~W.}\ \bibnamefont {Murch}},\ }\bibfield
  {title} {\bibinfo {title} {Quantum state tomography across the exceptional
  point in a single dissipative qubit},\ }\href
  {https://doi.org/10.1038/s41567-019-0652-z} {\bibfield  {journal} {\bibinfo
  {journal} {Nature Physics}\ }\textbf {\bibinfo {volume} {15}},\ \bibinfo
  {pages} {1232} (\bibinfo {year} {2019})}\BibitemShut {NoStop}%
\bibitem [{\citenamefont {Bergholtz}\ \emph {et~al.}(2021)\citenamefont
  {Bergholtz}, \citenamefont {Budich},\ and\ \citenamefont
  {Kunst}}]{Bergholtz_2021}%
  \BibitemOpen
  \bibfield  {author} {\bibinfo {author} {\bibfnamefont {E.~J.}\ \bibnamefont
  {Bergholtz}}, \bibinfo {author} {\bibfnamefont {J.~C.}\ \bibnamefont
  {Budich}},\ and\ \bibinfo {author} {\bibfnamefont {F.~K.}\ \bibnamefont
  {Kunst}},\ }\bibfield  {title} {\bibinfo {title} {Exceptional topology of
  non-hermitian systems},\ }\href
  {https://doi.org/10.1103/RevModPhys.93.015005} {\bibfield  {journal}
  {\bibinfo  {journal} {Rev. Mod. Phys.}\ }\textbf {\bibinfo {volume} {93}},\
  \bibinfo {pages} {015005} (\bibinfo {year} {2021})}\BibitemShut {NoStop}%
\bibitem [{\citenamefont {{\"O}zdemir}\ \emph {et~al.}(2019)\citenamefont
  {{\"O}zdemir}, \citenamefont {Rotter}, \citenamefont {Nori},\ and\
  \citenamefont {Yang}}]{Ozdemir_2019}%
  \BibitemOpen
  \bibfield  {author} {\bibinfo {author} {\bibfnamefont {{\c{S}}.~K.}\
  \bibnamefont {{\"O}zdemir}}, \bibinfo {author} {\bibfnamefont
  {S.}~\bibnamefont {Rotter}}, \bibinfo {author} {\bibfnamefont
  {F.}~\bibnamefont {Nori}},\ and\ \bibinfo {author} {\bibfnamefont
  {L.}~\bibnamefont {Yang}},\ }\bibfield  {title} {\bibinfo {title}
  {Parity--time symmetry and exceptional points in photonics},\ }\href
  {https://doi.org/10.1038/s41563-019-0304-9} {\bibfield  {journal} {\bibinfo
  {journal} {Nature Materials}\ }\textbf {\bibinfo {volume} {18}},\ \bibinfo
  {pages} {783} (\bibinfo {year} {2019})}\BibitemShut {NoStop}%
\bibitem [{\citenamefont {El-Ganainy}\ \emph {et~al.}(2018)\citenamefont
  {El-Ganainy}, \citenamefont {Makris}, \citenamefont {Khajavikhan},
  \citenamefont {Musslimani}, \citenamefont {Rotter},\ and\ \citenamefont
  {Christodoulides}}]{El_Ganainy_2018}%
  \BibitemOpen
  \bibfield  {author} {\bibinfo {author} {\bibfnamefont {R.}~\bibnamefont
  {El-Ganainy}}, \bibinfo {author} {\bibfnamefont {K.~G.}\ \bibnamefont
  {Makris}}, \bibinfo {author} {\bibfnamefont {M.}~\bibnamefont {Khajavikhan}},
  \bibinfo {author} {\bibfnamefont {Z.~H.}\ \bibnamefont {Musslimani}},
  \bibinfo {author} {\bibfnamefont {S.}~\bibnamefont {Rotter}},\ and\ \bibinfo
  {author} {\bibfnamefont {D.~N.}\ \bibnamefont {Christodoulides}},\ }\bibfield
   {title} {\bibinfo {title} {Non-hermitian physics and pt symmetry},\ }\href
  {https://doi.org/10.1038/nphys4323} {\bibfield  {journal} {\bibinfo
  {journal} {Nature Physics}\ }\textbf {\bibinfo {volume} {14}},\ \bibinfo
  {pages} {11} (\bibinfo {year} {2018})}\BibitemShut {NoStop}%
\bibitem [{\citenamefont {Feng}\ \emph {et~al.}(2017)\citenamefont {Feng},
  \citenamefont {El-Ganainy},\ and\ \citenamefont {Ge}}]{Feng_2017}%
  \BibitemOpen
  \bibfield  {author} {\bibinfo {author} {\bibfnamefont {L.}~\bibnamefont
  {Feng}}, \bibinfo {author} {\bibfnamefont {R.}~\bibnamefont {El-Ganainy}},\
  and\ \bibinfo {author} {\bibfnamefont {L.}~\bibnamefont {Ge}},\ }\bibfield
  {title} {\bibinfo {title} {Non-hermitian photonics based on parity--time
  symmetry},\ }\href {https://doi.org/10.1038/s41566-017-0031-1} {\bibfield
  {journal} {\bibinfo  {journal} {Nature Photonics}\ }\textbf {\bibinfo
  {volume} {11}},\ \bibinfo {pages} {752} (\bibinfo {year} {2017})}\BibitemShut
  {NoStop}%
\bibitem [{\citenamefont {Longhi}(2018)}]{Longhi_2017}%
  \BibitemOpen
  \bibfield  {author} {\bibinfo {author} {\bibfnamefont {S.}~\bibnamefont
  {Longhi}},\ }\bibfield  {title} {\bibinfo {title} {Parity-time symmetry meets
  photonics: A new twist in non-hermitian optics},\ }\href
  {https://doi.org/10.1209/0295-5075/120/64001} {\bibfield  {journal} {\bibinfo
   {journal} {Europhysics Letters}\ }\textbf {\bibinfo {volume} {120}},\
  \bibinfo {pages} {64001} (\bibinfo {year} {2018})}\BibitemShut {NoStop}%
\bibitem [{\citenamefont {Konotop}\ \emph {et~al.}(2016)\citenamefont
  {Konotop}, \citenamefont {Yang},\ and\ \citenamefont
  {Zezyulin}}]{Konotop_2016}%
  \BibitemOpen
  \bibfield  {author} {\bibinfo {author} {\bibfnamefont {V.~V.}\ \bibnamefont
  {Konotop}}, \bibinfo {author} {\bibfnamefont {J.}~\bibnamefont {Yang}},\ and\
  \bibinfo {author} {\bibfnamefont {D.~A.}\ \bibnamefont {Zezyulin}},\
  }\bibfield  {title} {\bibinfo {title} {Nonlinear waves in
  $\mathcal{PT}$-symmetric systems},\ }\href
  {https://doi.org/10.1103/RevModPhys.88.035002} {\bibfield  {journal}
  {\bibinfo  {journal} {Rev. Mod. Phys.}\ }\textbf {\bibinfo {volume} {88}},\
  \bibinfo {pages} {035002} (\bibinfo {year} {2016})}\BibitemShut {NoStop}%
\bibitem [{\citenamefont {Yuto~Ashida}\ and\ \citenamefont
  {Ueda}(2020)}]{non-hermitian-review}%
  \BibitemOpen
  \bibfield  {author} {\bibinfo {author} {\bibfnamefont {Z.~G.}\ \bibnamefont
  {Yuto~Ashida}}\ and\ \bibinfo {author} {\bibfnamefont {M.}~\bibnamefont
  {Ueda}},\ }\bibfield  {title} {\bibinfo {title} {Non-hermitian physics},\
  }\href {https://doi.org/10.1080/00018732.2021.1876991} {\bibfield  {journal}
  {\bibinfo  {journal} {Advances in Physics}\ }\textbf {\bibinfo {volume}
  {69}},\ \bibinfo {pages} {249} (\bibinfo {year} {2020})}\BibitemShut
  {NoStop}%
\bibitem [{\citenamefont {Gohsrich}\ \emph {et~al.}(2024)\citenamefont
  {Gohsrich}, \citenamefont {Fauman},\ and\ \citenamefont
  {Kunst}}]{general_hatano_nelson}%
  \BibitemOpen
  \bibfield  {author} {\bibinfo {author} {\bibfnamefont {J.~T.}\ \bibnamefont
  {Gohsrich}}, \bibinfo {author} {\bibfnamefont {J.}~\bibnamefont {Fauman}},\
  and\ \bibinfo {author} {\bibfnamefont {F.~K.}\ \bibnamefont {Kunst}},\ }\href
  {https://arxiv.org/abs/2403.12018} {\bibinfo {title} {Exceptional points of
  any order in a generalized hatano-nelson model}} (\bibinfo {year} {2024}),\
  \Eprint {https://arxiv.org/abs/2403.12018} {arXiv:2403.12018} \BibitemShut
  {NoStop}%
\bibitem [{\citenamefont {Wiersig}(2022)}]{higher_order_exceptional}%
  \BibitemOpen
  \bibfield  {author} {\bibinfo {author} {\bibfnamefont {J.}~\bibnamefont
  {Wiersig}},\ }\bibfield  {title} {\bibinfo {title} {Revisiting the
  hierarchical construction of higher-order exceptional points},\ }\href
  {https://doi.org/10.1103/PhysRevA.106.063526} {\bibfield  {journal} {\bibinfo
   {journal} {Phys. Rev. A}\ }\textbf {\bibinfo {volume} {106}},\ \bibinfo
  {pages} {063526} (\bibinfo {year} {2022})}\BibitemShut {NoStop}%
\bibitem [{\citenamefont {Hashemi}\ \emph {et~al.}(2022)\citenamefont
  {Hashemi}, \citenamefont {Busch}, \citenamefont {Christodoulides},
  \citenamefont {Ozdemir},\ and\ \citenamefont
  {El-Ganainy}}]{open_exceptional}%
  \BibitemOpen
  \bibfield  {author} {\bibinfo {author} {\bibfnamefont {A.}~\bibnamefont
  {Hashemi}}, \bibinfo {author} {\bibfnamefont {K.}~\bibnamefont {Busch}},
  \bibinfo {author} {\bibfnamefont {D.~N.}\ \bibnamefont {Christodoulides}},
  \bibinfo {author} {\bibfnamefont {S.~K.}\ \bibnamefont {Ozdemir}},\ and\
  \bibinfo {author} {\bibfnamefont {R.}~\bibnamefont {El-Ganainy}},\ }\bibfield
   {title} {\bibinfo {title} {Linear response theory of open systems with
  exceptional points},\ }\href {https://doi.org/10.1038/s41467-022-30715-8}
  {\bibfield  {journal} {\bibinfo  {journal} {Nature Communications}\ }\textbf
  {\bibinfo {volume} {13}},\ \bibinfo {pages} {3281} (\bibinfo {year}
  {2022})}\BibitemShut {NoStop}%
\bibitem [{\citenamefont {Ding}\ \emph {et~al.}(2022)\citenamefont {Ding},
  \citenamefont {Fang},\ and\ \citenamefont {Ma}}]{toplology_non_hermitian}%
  \BibitemOpen
  \bibfield  {author} {\bibinfo {author} {\bibfnamefont {K.}~\bibnamefont
  {Ding}}, \bibinfo {author} {\bibfnamefont {C.}~\bibnamefont {Fang}},\ and\
  \bibinfo {author} {\bibfnamefont {G.}~\bibnamefont {Ma}},\ }\bibfield
  {title} {\bibinfo {title} {Non-hermitian topology and exceptional-point
  geometries},\ }\href {https://doi.org/10.1038/s42254-022-00516-5} {\bibfield
  {journal} {\bibinfo  {journal} {Nature Reviews Physics}\ }\textbf {\bibinfo
  {volume} {4}},\ \bibinfo {pages} {745} (\bibinfo {year} {2022})}\BibitemShut
  {NoStop}%
\bibitem [{\citenamefont {Othman}\ \emph {et~al.}(2017)\citenamefont {Othman},
  \citenamefont {Galdi},\ and\ \citenamefont {Capolino}}]{EPD_photonic}%
  \BibitemOpen
  \bibfield  {author} {\bibinfo {author} {\bibfnamefont {M.~A.~K.}\
  \bibnamefont {Othman}}, \bibinfo {author} {\bibfnamefont {V.}~\bibnamefont
  {Galdi}},\ and\ \bibinfo {author} {\bibfnamefont {F.}~\bibnamefont
  {Capolino}},\ }\bibfield  {title} {\bibinfo {title} {Exceptional points of
  degeneracy and $\mathcal{P}\mathcal{T}$ symmetry in photonic coupled chains
  of scatterers},\ }\href {https://doi.org/10.1103/PhysRevB.95.104305}
  {\bibfield  {journal} {\bibinfo  {journal} {Phys. Rev. B}\ }\textbf {\bibinfo
  {volume} {95}},\ \bibinfo {pages} {104305} (\bibinfo {year}
  {2017})}\BibitemShut {NoStop}%
\bibitem [{\citenamefont {Mealy}\ and\ \citenamefont
  {Capolino}(2023)}]{wave_guideEPD}%
  \BibitemOpen
  \bibfield  {author} {\bibinfo {author} {\bibfnamefont {T.}~\bibnamefont
  {Mealy}}\ and\ \bibinfo {author} {\bibfnamefont {F.}~\bibnamefont
  {Capolino}},\ }\bibfield  {title} {\bibinfo {title} {Exceptional points of
  degeneracy with indirect band gap induced by mixing forward and backward
  propagating waves},\ }\href {https://doi.org/10.1103/PhysRevA.107.012214}
  {\bibfield  {journal} {\bibinfo  {journal} {Phys. Rev. A}\ }\textbf {\bibinfo
  {volume} {107}},\ \bibinfo {pages} {012214} (\bibinfo {year}
  {2023})}\BibitemShut {NoStop}%
\bibitem [{\citenamefont {Mandal}\ and\ \citenamefont
  {Bergholtz}(2021)}]{symmetry_higher_order_EP}%
  \BibitemOpen
  \bibfield  {author} {\bibinfo {author} {\bibfnamefont {I.}~\bibnamefont
  {Mandal}}\ and\ \bibinfo {author} {\bibfnamefont {E.~J.}\ \bibnamefont
  {Bergholtz}},\ }\bibfield  {title} {\bibinfo {title} {Symmetry and
  higher-order exceptional points},\ }\href
  {https://doi.org/10.1103/PhysRevLett.127.186601} {\bibfield  {journal}
  {\bibinfo  {journal} {Phys. Rev. Lett.}\ }\textbf {\bibinfo {volume} {127}},\
  \bibinfo {pages} {186601} (\bibinfo {year} {2021})}\BibitemShut {NoStop}%
\bibitem [{\citenamefont {Saha}\ \emph
  {et~al.}(2023{\natexlab{a}})\citenamefont {Saha}, \citenamefont {Kulkarni},\
  and\ \citenamefont {Agarwalla}}]{finite-range-subdiffusive}%
  \BibitemOpen
  \bibfield  {author} {\bibinfo {author} {\bibfnamefont {M.}~\bibnamefont
  {Saha}}, \bibinfo {author} {\bibfnamefont {M.}~\bibnamefont {Kulkarni}},\
  and\ \bibinfo {author} {\bibfnamefont {B.~K.}\ \bibnamefont {Agarwalla}},\
  }\bibfield  {title} {\bibinfo {title} {Exceptional hypersurfaces of transfer
  matrices of finite-range lattice models and their consequences on quantum
  transport properties},\ }\href {https://doi.org/10.1103/PhysRevB.108.075406}
  {\bibfield  {journal} {\bibinfo  {journal} {Phys. Rev. B}\ }\textbf {\bibinfo
  {volume} {108}},\ \bibinfo {pages} {075406} (\bibinfo {year}
  {2023}{\natexlab{a}})}\BibitemShut {NoStop}%
\bibitem [{\citenamefont {Saha}\ \emph {et~al.}(2024)\citenamefont {Saha},
  \citenamefont {Agarwalla}, \citenamefont {Kulkarni},\ and\ \citenamefont
  {Purkayastha}}]{saha2024effect}%
  \BibitemOpen
  \bibfield  {author} {\bibinfo {author} {\bibfnamefont {M.}~\bibnamefont
  {Saha}}, \bibinfo {author} {\bibfnamefont {B.~K.}\ \bibnamefont {Agarwalla}},
  \bibinfo {author} {\bibfnamefont {M.}~\bibnamefont {Kulkarni}},\ and\
  \bibinfo {author} {\bibfnamefont {A.}~\bibnamefont {Purkayastha}},\
  }\bibfield  {title} {\bibinfo {title} {Effect of order of transfer matrix
  exceptional points on transport at band edges},\ }\href
  {https://arxiv.org/abs/2407.10884} {\bibfield  {journal} {\bibinfo  {journal}
  {arXiv preprint arXiv:2407.10884}\ } (\bibinfo {year} {2024})}\BibitemShut
  {NoStop}%
\bibitem [{\citenamefont {Zhong}\ \emph {et~al.}(2019)\citenamefont {Zhong},
  \citenamefont {Ren}, \citenamefont {Khajavikhan}, \citenamefont
  {Christodoulides}, \citenamefont {\"Ozdemir},\ and\ \citenamefont
  {El-Ganainy}}]{sensing_EPD}%
  \BibitemOpen
  \bibfield  {author} {\bibinfo {author} {\bibfnamefont {Q.}~\bibnamefont
  {Zhong}}, \bibinfo {author} {\bibfnamefont {J.}~\bibnamefont {Ren}}, \bibinfo
  {author} {\bibfnamefont {M.}~\bibnamefont {Khajavikhan}}, \bibinfo {author}
  {\bibfnamefont {D.~N.}\ \bibnamefont {Christodoulides}}, \bibinfo {author}
  {\bibfnamefont {i.~m. c.~K.}\ \bibnamefont {\"Ozdemir}},\ and\ \bibinfo
  {author} {\bibfnamefont {R.}~\bibnamefont {El-Ganainy}},\ }\bibfield  {title}
  {\bibinfo {title} {Sensing with exceptional surfaces in order to combine
  sensitivity with robustness},\ }\href
  {https://doi.org/10.1103/PhysRevLett.122.153902} {\bibfield  {journal}
  {\bibinfo  {journal} {Phys. Rev. Lett.}\ }\textbf {\bibinfo {volume} {122}},\
  \bibinfo {pages} {153902} (\bibinfo {year} {2019})}\BibitemShut {NoStop}%
\bibitem [{\citenamefont {McDonald}\ and\ \citenamefont
  {Clerk}(2020)}]{sensing1}%
  \BibitemOpen
  \bibfield  {author} {\bibinfo {author} {\bibfnamefont {A.}~\bibnamefont
  {McDonald}}\ and\ \bibinfo {author} {\bibfnamefont {A.~A.}\ \bibnamefont
  {Clerk}},\ }\bibfield  {title} {\bibinfo {title} {Exponentially-enhanced
  quantum sensing with non-hermitian lattice dynamics},\ }\href
  {https://doi.org/10.1038/s41467-020-19090-4} {\bibfield  {journal} {\bibinfo
  {journal} {Nature Communications}\ }\textbf {\bibinfo {volume} {11}},\
  \bibinfo {pages} {5382} (\bibinfo {year} {2020})}\BibitemShut {NoStop}%
\bibitem [{\citenamefont {De~Carlo}\ \emph {et~al.}(2022)\citenamefont
  {De~Carlo}, \citenamefont {De~Leonardis}, \citenamefont {Soref},
  \citenamefont {Colatorti},\ and\ \citenamefont {Passaro}}]{sensing_review}%
  \BibitemOpen
  \bibfield  {author} {\bibinfo {author} {\bibfnamefont {M.}~\bibnamefont
  {De~Carlo}}, \bibinfo {author} {\bibfnamefont {F.}~\bibnamefont
  {De~Leonardis}}, \bibinfo {author} {\bibfnamefont {R.~A.}\ \bibnamefont
  {Soref}}, \bibinfo {author} {\bibfnamefont {L.}~\bibnamefont {Colatorti}},\
  and\ \bibinfo {author} {\bibfnamefont {V.~M.~N.}\ \bibnamefont {Passaro}},\
  }\bibfield  {title} {\bibinfo {title} {Non-hermitian sensing in photonics and
  electronics: A review},\ }\href {https://www.mdpi.com/1424-8220/22/11/3977}
  {\bibfield  {journal} {\bibinfo  {journal} {Sensors}\ }\textbf {\bibinfo
  {volume} {22}} (\bibinfo {year} {2022})}\BibitemShut {NoStop}%
\bibitem [{\citenamefont {Wiersig}(2020)}]{sensing_review1}%
  \BibitemOpen
  \bibfield  {author} {\bibinfo {author} {\bibfnamefont {J.}~\bibnamefont
  {Wiersig}},\ }\bibfield  {title} {\bibinfo {title} {Review of exceptional
  point-based sensors},\ }\href {https://doi.org/10.1364/PRJ.396115} {\bibfield
   {journal} {\bibinfo  {journal} {Photon. Res.}\ }\textbf {\bibinfo {volume}
  {8}},\ \bibinfo {pages} {1457} (\bibinfo {year} {2020})}\BibitemShut
  {NoStop}%
\bibitem [{\citenamefont {Veysi}\ \emph {et~al.}(2018)\citenamefont {Veysi},
  \citenamefont {Othman}, \citenamefont {Figotin},\ and\ \citenamefont
  {Capolino}}]{band_edge_EP_transfer_matrix}%
  \BibitemOpen
  \bibfield  {author} {\bibinfo {author} {\bibfnamefont {M.}~\bibnamefont
  {Veysi}}, \bibinfo {author} {\bibfnamefont {M.~A.~K.}\ \bibnamefont
  {Othman}}, \bibinfo {author} {\bibfnamefont {A.}~\bibnamefont {Figotin}},\
  and\ \bibinfo {author} {\bibfnamefont {F.}~\bibnamefont {Capolino}},\
  }\bibfield  {title} {\bibinfo {title} {Degenerate band edge laser},\ }\href
  {https://doi.org/10.1103/PhysRevB.97.195107} {\bibfield  {journal} {\bibinfo
  {journal} {Phys. Rev. B}\ }\textbf {\bibinfo {volume} {97}},\ \bibinfo
  {pages} {195107} (\bibinfo {year} {2018})}\BibitemShut {NoStop}%
\bibitem [{\citenamefont {Zhong}\ \emph {et~al.}(2021)\citenamefont {Zhong},
  \citenamefont {Hashemi}, \citenamefont {\"Ozdemir},\ and\ \citenamefont
  {El-Ganainy}}]{spontaneous_emission}%
  \BibitemOpen
  \bibfield  {author} {\bibinfo {author} {\bibfnamefont {Q.}~\bibnamefont
  {Zhong}}, \bibinfo {author} {\bibfnamefont {A.}~\bibnamefont {Hashemi}},
  \bibinfo {author} {\bibfnamefont {i.~m. c.~K.}\ \bibnamefont {\"Ozdemir}},\
  and\ \bibinfo {author} {\bibfnamefont {R.}~\bibnamefont {El-Ganainy}},\
  }\bibfield  {title} {\bibinfo {title} {Control of spontaneous emission
  dynamics in microcavities with chiral exceptional surfaces},\ }\href
  {https://doi.org/10.1103/PhysRevResearch.3.013220} {\bibfield  {journal}
  {\bibinfo  {journal} {Phys. Rev. Res.}\ }\textbf {\bibinfo {volume} {3}},\
  \bibinfo {pages} {013220} (\bibinfo {year} {2021})}\BibitemShut {NoStop}%
\bibitem [{\citenamefont {Soleymani}\ \emph {et~al.}(2022)\citenamefont
  {Soleymani}, \citenamefont {Zhong}, \citenamefont {Mokim}, \citenamefont
  {Rotter}, \citenamefont {El-Ganainy},\ and\ \citenamefont
  {{\"O}zdemir}}]{perfect_absorber}%
  \BibitemOpen
  \bibfield  {author} {\bibinfo {author} {\bibfnamefont {S.}~\bibnamefont
  {Soleymani}}, \bibinfo {author} {\bibfnamefont {Q.}~\bibnamefont {Zhong}},
  \bibinfo {author} {\bibfnamefont {M.}~\bibnamefont {Mokim}}, \bibinfo
  {author} {\bibfnamefont {S.}~\bibnamefont {Rotter}}, \bibinfo {author}
  {\bibfnamefont {R.}~\bibnamefont {El-Ganainy}},\ and\ \bibinfo {author}
  {\bibfnamefont {{\c S}.~K.}\ \bibnamefont {{\"O}zdemir}},\ }\bibfield
  {title} {\bibinfo {title} {Chiral and degenerate perfect absorption on
  exceptional surfaces},\ }\href {https://doi.org/10.1038/s41467-022-27990-w}
  {\bibfield  {journal} {\bibinfo  {journal} {Nature Communications}\ }\textbf
  {\bibinfo {volume} {13}},\ \bibinfo {pages} {599} (\bibinfo {year}
  {2022})}\BibitemShut {NoStop}%
\bibitem [{\citenamefont {Lavis}\ and\ \citenamefont
  {Southern}(1997)}]{matrix_inversion}%
  \BibitemOpen
  \bibfield  {author} {\bibinfo {author} {\bibfnamefont {D.}~\bibnamefont
  {Lavis}}\ and\ \bibinfo {author} {\bibfnamefont {B.}~\bibnamefont
  {Southern}},\ }\bibfield  {title} {\bibinfo {title} {The inverse of a
  symmetric banded toeplitz matrix},\ }\href
  {https://doi.org/https://doi.org/10.1016/S0034-4877(97)81478-4} {\bibfield
  {journal} {\bibinfo  {journal} {Reports on Mathematical Physics}\ }\textbf
  {\bibinfo {volume} {39}},\ \bibinfo {pages} {137} (\bibinfo {year}
  {1997})}\BibitemShut {NoStop}%
\bibitem [{\citenamefont {Molinari}(1997)}]{Molinari_1997}%
  \BibitemOpen
  \bibfield  {author} {\bibinfo {author} {\bibfnamefont {L.}~\bibnamefont
  {Molinari}},\ }\bibfield  {title} {\bibinfo {title} {Transfer matrices and
  tridiagonal-block hamiltonians with periodic and scattering boundary
  conditions},\ }\href {https://doi.org/10.1088/0305-4470/30/3/021} {\bibfield
  {journal} {\bibinfo  {journal} {Journal of Physics A: Mathematical and
  General}\ }\textbf {\bibinfo {volume} {30}},\ \bibinfo {pages} {983}
  (\bibinfo {year} {1997})}\BibitemShut {NoStop}%
\bibitem [{\citenamefont {Molinari}(1998)}]{Molinari_1998}%
  \BibitemOpen
  \bibfield  {author} {\bibinfo {author} {\bibfnamefont {L.}~\bibnamefont
  {Molinari}},\ }\bibfield  {title} {\bibinfo {title} {Transfer matrices,
  non-hermitian hamiltonians and resolvents: some spectral identities},\ }\href
  {https://doi.org/10.1088/0305-4470/31/42/014} {\bibfield  {journal} {\bibinfo
   {journal} {Journal of Physics A: Mathematical and General}\ }\textbf
  {\bibinfo {volume} {31}},\ \bibinfo {pages} {8553} (\bibinfo {year}
  {1998})}\BibitemShut {NoStop}%
\bibitem [{\citenamefont {Dwivedi}\ and\ \citenamefont
  {Chua}(2016)}]{transfer_matrix1}%
  \BibitemOpen
  \bibfield  {author} {\bibinfo {author} {\bibfnamefont {V.}~\bibnamefont
  {Dwivedi}}\ and\ \bibinfo {author} {\bibfnamefont {V.}~\bibnamefont {Chua}},\
  }\bibfield  {title} {\bibinfo {title} {Of bulk and boundaries: Generalized
  transfer matrices for tight-binding models},\ }\href
  {https://doi.org/10.1103/PhysRevB.93.134304} {\bibfield  {journal} {\bibinfo
  {journal} {Phys. Rev. B}\ }\textbf {\bibinfo {volume} {93}},\ \bibinfo
  {pages} {134304} (\bibinfo {year} {2016})}\BibitemShut {NoStop}%
\bibitem [{\citenamefont {Kunst}\ and\ \citenamefont
  {Dwivedi}(2019)}]{transfer_matrix2}%
  \BibitemOpen
  \bibfield  {author} {\bibinfo {author} {\bibfnamefont {F.~K.}\ \bibnamefont
  {Kunst}}\ and\ \bibinfo {author} {\bibfnamefont {V.}~\bibnamefont
  {Dwivedi}},\ }\bibfield  {title} {\bibinfo {title} {Non-hermitian systems and
  topology: A transfer-matrix perspective},\ }\href
  {https://doi.org/10.1103/PhysRevB.99.245116} {\bibfield  {journal} {\bibinfo
  {journal} {Phys. Rev. B}\ }\textbf {\bibinfo {volume} {99}},\ \bibinfo
  {pages} {245116} (\bibinfo {year} {2019})}\BibitemShut {NoStop}%
\bibitem [{\citenamefont {Saha}\ \emph
  {et~al.}(2023{\natexlab{b}})\citenamefont {Saha}, \citenamefont {Agarwalla},
  \citenamefont {Kulkarni},\ and\ \citenamefont
  {Purkayastha}}]{universal_subdiffusive}%
  \BibitemOpen
  \bibfield  {author} {\bibinfo {author} {\bibfnamefont {M.}~\bibnamefont
  {Saha}}, \bibinfo {author} {\bibfnamefont {B.~K.}\ \bibnamefont {Agarwalla}},
  \bibinfo {author} {\bibfnamefont {M.}~\bibnamefont {Kulkarni}},\ and\
  \bibinfo {author} {\bibfnamefont {A.}~\bibnamefont {Purkayastha}},\
  }\bibfield  {title} {\bibinfo {title} {Universal subdiffusive behavior at
  band edges from transfer matrix exceptional points},\ }\href
  {https://doi.org/10.1103/PhysRevLett.130.187101} {\bibfield  {journal}
  {\bibinfo  {journal} {Phys. Rev. Lett.}\ }\textbf {\bibinfo {volume} {130}},\
  \bibinfo {pages} {187101} (\bibinfo {year} {2023}{\natexlab{b}})}\BibitemShut
  {NoStop}%
\bibitem [{\citenamefont {Saha}\ \emph
  {et~al.}(2023{\natexlab{c}})\citenamefont {Saha}, \citenamefont {Agarwalla},
  \citenamefont {Kulkarni},\ and\ \citenamefont
  {Purkayastha}}]{superballistic1}%
  \BibitemOpen
  \bibfield  {author} {\bibinfo {author} {\bibfnamefont {M.}~\bibnamefont
  {Saha}}, \bibinfo {author} {\bibfnamefont {B.~K.}\ \bibnamefont {Agarwalla}},
  \bibinfo {author} {\bibfnamefont {M.}~\bibnamefont {Kulkarni}},\ and\
  \bibinfo {author} {\bibfnamefont {A.}~\bibnamefont {Purkayastha}},\
  }\bibfield  {title} {\bibinfo {title} {Environment assisted superballistic
  scaling of conductance},\ }\href
  {https://doi.org/10.1103/PhysRevB.108.L161115} {\bibfield  {journal}
  {\bibinfo  {journal} {Phys. Rev. B}\ }\textbf {\bibinfo {volume} {108}},\
  \bibinfo {pages} {L161115} (\bibinfo {year}
  {2023}{\natexlab{c}})}\BibitemShut {NoStop}%
\bibitem [{\citenamefont {Hu}\ and\ \citenamefont {Zhang}(2024)}]{hu_2024}%
  \BibitemOpen
  \bibfield  {author} {\bibinfo {author} {\bibfnamefont {X.-D.}\ \bibnamefont
  {Hu}}\ and\ \bibinfo {author} {\bibfnamefont {D.-B.}\ \bibnamefont {Zhang}},\
  }\bibfield  {title} {\bibinfo {title} {Exact correlation functions for
  dual-unitary quantum circuits with exceptional points},\ }\href
  {https://arxiv.org/abs/2406.08338} {\bibfield  {journal} {\bibinfo  {journal}
  {arXiv:2406.08338}\ } (\bibinfo {year} {2024})}\BibitemShut {NoStop}%
\end{thebibliography}%
\end{document}